\documentclass[11pt,a4paper]{article}
\input amssymb.sty
\begin{document}


\newcounter{mo}
\newcommand{\mo}[1]
{\stepcounter{mo}$^{\bf MO\themo}$%
\footnotetext{\hspace{-3.7mm}$^{\blacksquare\!\blacksquare}$
{\bf MO\themo:~}#1}}

\newcounter{bk}
\newcommand{\bk}[1]
{\stepcounter{bk}$^{\bf BK\thebk}$%
\footnotetext{\hspace{-3.7mm}$^{\blacksquare\!\blacksquare}$
{\bf BK\thebk:~}#1}}


\newcommand{\Si}{\Sigma}
\newcommand{\tr}{{\rm tr}}
\newcommand{\ad}{{\rm ad}}
\newcommand{\Ad}{{\rm Ad}}
\newcommand{\ti}[1]{\tilde{#1}}
\newcommand{\om}{\omega}
\newcommand{\Om}{\Omega}
\newcommand{\de}{\delta}
\newcommand{\al}{\alpha}
\newcommand{\te}{\theta}
\newcommand{\vth}{\vartheta}
\newcommand{\be}{\beta}
\newcommand{\la}{\lambda}
\newcommand{\La}{\Lambda}
\newcommand{\D}{\Delta}
\newcommand{\ve}{\varepsilon}
\newcommand{\ep}{\epsilon}
\newcommand{\vf}{\varphi}
\newcommand{\vfh}{\varphi^\hbar}
\newcommand{\vfe}{\varphi^\eta}
\newcommand{\fh}{\phi^\hbar}
\newcommand{\fe}{\phi^\eta}
\newcommand{\G}{\Gamma}
\newcommand{\ka}{\kappa}
\newcommand{\ip}{\hat{\upsilon}}
\newcommand{\Ip}{\hat{\Upsilon}}
\newcommand{\ga}{\gamma}
\newcommand{\ze}{\zeta}
\newcommand{\si}{\sigma}

\def\hS{{\hat{S}}}

\newcommand{\li}{\lim_{n\rightarrow \infty}}
\def\mapright#1{\smash{
\mathop{\longrightarrow}\limits^{#1}}}

\newcommand{\mat}[4]{\left(\begin{array}{cc}{#1}&{#2}\\{#3}&{#4}
\end{array}\right)}
\newcommand{\thmat}[9]{\left(
\begin{array}{ccc}{#1}&{#2}&{#3}\\{#4}&{#5}&{#6}\\
{#7}&{#8}&{#9}
\end{array}\right)}
\newcommand{\beq}[1]{\begin{equation}\label{#1}}
\newcommand{\eq}{\end{equation}}
\newcommand{\beqn}[1]{\begin{eqnarray}\label{#1}}
\newcommand{\eqn}{\end{eqnarray}}
\newcommand{\p}{\partial}
\def\sq2{\sqrt{2}}
\newcommand{\di}{{\rm diag}}
\newcommand{\oh}{\frac{1}{2}}
\newcommand{\su}{{\bf su_2}}
\newcommand{\sun}{{\bf su_n}}
\newcommand{\uo}{{\bf u_1}}
\newcommand{\SL}{{\rm SL}(2,{\mathbb C})}
\newcommand{\GLN}{{\rm GL}(N,{\mathbb C})}
\newcommand{\SUN}{{\rm SU}(N)}
\def\sln{{\rm sl}(N, {\mathbb C})}
\def\sl2{{\rm sl}(2, {\mathbb C})}
\def\SLN{{\rm SL}(N, {\mathbb C})}
\def\SLT{{\rm SL}(2, {\mathbb C})}
\def\PSLN{{\rm PSL}(N, {\mathbb C})}
\newcommand{\PGLN}{{\rm PGL}(N,{\mathbb C})}
\newcommand{\gln}{{\rm gl}(N, {\mathbb C})}
\newcommand{\PSL}{{\rm PSL}_2( {\mathbb Z})}
\def\f1#1{\frac{1}{#1}}
\def\lb{\lfloor}
\def\rb{\rfloor}
\def\sn{{\rm sn}}
\def\cn{{\rm cn}}
\def\dn{{\rm dn}}
\newcommand{\rar}{\rightarrow}
\newcommand{\upar}{\uparrow}
\newcommand{\sm}{\setminus}
\newcommand{\ms}{\mapsto}
\newcommand{\bp}{\bar{\partial}}
\newcommand{\bz}{\bar{z}}
\newcommand{\bw}{\bar{w}}
\newcommand{\bA}{\bar{A}}
\newcommand{\bG}{\bar{G}}
\newcommand{\bL}{\bar{L}}
\newcommand{\btau}{\bar{\tau}}

\newcommand{\tie}{\tilde{e}}
\newcommand{\tial}{\tilde{\alpha}}

\newcommand{\Sh}{\hat{S}}
\newcommand{\vtb}{\theta_{2}}
\newcommand{\vtc}{\theta_{3}}
\newcommand{\vtd}{\theta_{4}}

\def\mC{{\mathbb C}}
\def\mZ{{\mathbb Z}}
\def\mR{{\mathbb R}}
\def\mN{{\mathbb N}}

\def\frak{\mathfrak}
\def\gg{{\frak g}}
\def\gJ{{\frak J}}
\def\gS{{\frak S}}
\def\gL{{\frak L}}
\def\gG{{\frak G}}
\def\gB{{\frak B}}
\def\gk{{\frak k}}
\def\gK{{\frak K}}
\def\gl{{\frak l}}
\def\gh{{\frak h}}
\def\gH{{\frak H}}
\def\gt{{\frak t}}
\def\gT{{\frak T}}

\def\baal{\bar{\al}}
\def\babe{\bar{\be}}

\def\bfA{{\bf A}}
\def\bfa{{\bf a}}
\def\bfb{{\bf b}}
\def\bfc{{\bf c}}
\def\bfd{{\bf d}}
\def\bfe{{\bf e}}
\def\bff{{\bf f}}
\def\bfg{{\bf g}}
\def\bfm{{\bf m}}
\def\bfn{{\bf n}}
\def\bfp{{\bf p}}
\def\bfu{{\bf u}}
\def\bfv{{\bf v}}
\def\bfr{{\bf r}}
\def\bfs{{\bf s}}
\def\bft{{\bf t}}
\def\bfx{{\bf x}}
\def\bfM{{\bf M}}
\def\bfR{{\bf R}}
\def\bfC{{\bf C}}
\def\bfS{{\bf S}}
\def\bfJ{{\bf J}}
\def\bfz{{\bf z}}
\def\bfnu{{\bf \nu}}
\def\bfsi{{\bf \sigma}}
\def\bfU{{\bf U}}

\def\clA{\mathcal{A}}
\def\clC{\mathcal{C}}
\def\clD{\mathcal{D}}
\def\clE{\mathcal{E}}
\def\clF{\mathcal{F}}
\def\clG{\mathcal{G}}
\def\clR{\mathcal{R}}
\def\clU{\mathcal{U}}
\def\clT{\mathcal{T}}
\def\clO{\mathcal{O}}
\def\clH{\mathcal{H}}
\def\clK{\mathcal{K}}
\def\clJ{\mathcal{J}}
\def\clL{\mathcal{L}}
\def\clM{\mathcal{M}}
\def\clN{\mathcal{N}}
\def\clP{\mathcal{P}}
\def\clQ{\mathcal{Q}}
\def\clW{\mathcal{W}}
\def\clY{\mathcal{Y}}
\def\clZ{\mathcal{Z}}

\def\ba6{{\bf e_6}}
\def\bae7{{\bf e_7}}
\def\bag2{{\bf g_2}}
\def\baso8{{\bf so(8)}}
\def\baso{{\bf so(n)}}

\def\sr2{\sqrt{2}}
\newcommand{\ran}{\rangle}
\newcommand{\lan}{\langle}
\def\f1#1{\frac{1}{#1}}
\def\lb{\lfloor}
\def\rb{\rfloor}
\newcommand{\slim}[2]{\sum\limits_{#1}^{#2}}

\newcommand{\sect}[1]{\setcounter{equation}{0}\section{#1}}
\renewcommand{\theequation}{\thesection.\arabic{equation}}
\newtheorem{predl}{Proposition}[section]
\newtheorem{defi}{Definition}[section]
\newtheorem{rem}{Remark}[section]
\newtheorem{cor}{Corollary}[section]
\newtheorem{lem}{Lemma}[section]
\newtheorem{theor}{Theorem}[section]

\vspace{0.3in}
\begin{flushright}
 ITEP-TH-67/09\\
\end{flushright}
\vspace{10mm}
\begin{center}
{\Large{\bf Gauge theories, Simple Groups and Integrable Systems}
}\\
\vspace{5mm}
M.A.Olshanetsky,\\
{\sf Institute of Theoretical and Experimental Physics, Moscow, Russia,}\\
{\em  olshanet@itep.ru}\\
\vspace{5mm}
\end{center}

\begin{abstract}
In this review we discuss interrelations between classical Hitchin integrable systems,
monodromy preserving equations and topological field theories coming
from N=4 supersymmetric Yang-Mills theories developed by Gukov, Kapustin and Witten.
In particular, we define the systems related to bundles with nontrivial characteristic
classes and discuss  relations of the  characteristic
classes with monopole configurations in the Yang-Mills theory.
\end{abstract}


\tableofcontents

\section{Introduction}
\setcounter{equation}{0}
This brief review is my tribute to Lev Borisovich Okun.
He  inspired my interest to group-theoretical methods in
gauge theories.

There are many aspects of interrelations between finite-dimensional integrable systems and gauge theories. For quantum systems see \cite{NSh,GSh,BB} and references therein. Here we consider this connection in the classical case, which was described first in
 two fundamental papers of Hitchin \cite{Hi1,Hi2}. It was demonstrated there that   the  self-duality equation can be used to derive a wide class of finite-dimensional
classical integrable systems - the so-called Hitchin systems, as well as their
non-autonomous version - the monodromy preserving equations. We will consider
this construction here in details. Later it was shown
 that some well-known integrable systems can be derived in this way \cite{Ma,GN,ER,Ne,LO,Kr3,LOZ1,LOZ2}.

It was found recently that this construction appears in the reformulation of the geometric Langlands program in terms of  $N=4$ SUSY
Yang-Mills theory \cite{KW,GW,GW1}.
As a by product, it was demonstrated in \cite{KW} that some important equivalence
in integrable systems - the so-called the Symplectic Hecke Correspondence
\cite{LOZ1} can be
explained in terms of the t'Hooft operators. It be a subject of the last parts of the paper.

\bigskip
{\small {\bf Acknowledgments.}\\
 The work was supported by grants RFBR-09-02-00393, RFBR-09-01-92437-CE$_a$,
NSh-3036.2008.2.}


\section{Classical Integrable Systems}
\setcounter{equation}{0}

\subsection{Integrability}

We consider here the integrable systems of classical mechanics
\cite{Ar1}.
In this case the notion of complete integrability can be formulated correctly,
while in a field theory there are subtleties in its definition.

Consider a smooth symplectic manifold $\clR\,$ of $\dim(\mathcal{R})=2n$.
It means that there exists a closed non-degenerate  two-form $\om$, and the inverse
bivector $\pi\,$
$\,(\om_{a,b}\pi^{bc}=\de^c_a)$, such that the space of smooth functions $C^\infty(\clR)$ becomes a Poisson algebra with respect to the Poisson brackets
\beq{pb}
\{F,G\}=\lan dF|\pi| dG\ran=\p_aF\pi^{ab}\p_bG\equiv
\p_aF\pi^{ab}\p_bG\,.
\eq
In terms of the bi-vector $\pi^{ab}$  the Jacobi identity for the brackets assumes the form
$$
\pi^{ab}\p_b\pi^{cd}+\pi^{cb}\p_b\pi^{da}+\pi^{db}\p_b\pi^{ca}=0\,.
$$

\emph{Example 1}\\
For $\clR=\mR^{2n}=\{(p_1,\ldots,p_n,q_1,\ldots,q_n)\}$ define
$$
\pi=\left(
      \begin{array}{cc}
        0 & E \\
        -E & 0 \\
      \end{array}
    \right)\,.
$$
Then (\ref{pb}) is the Darboux brackets.
These brackets can be defined locally on any cotangent bundle $T^*M$
to a manifold $M\,$ $(\dim\, M=n)$, where $\{(p_1,\ldots,p_n,q_1,\ldots,q_n)\}$ are local coordinates.

\emph{Example 2}\\
Consider a Lie algebra $\gg$
and let $\{e_a\,,~a=1,\ldots n\}\,$ $\,(n=dim\,\gg)$ be the set of its generators
with the commutation relations
$$
[e_a,e_b]=C_{ab}^ce_c\,.
$$
Let $\gg^*$ be dual to $\gg$ space (the Lie coalgebra) with respect to
a pairing $\lan~,~\ran$. Define a basis $\{E^b\}$ in $\gg^*$  such that
$$
\lan e_a,E^b\ran=\de_a^b\,.
$$

Let $F(\bfS)$, $G(\bfS)$ $\,(\bfS=S_aE^a)$ be two functions on $\gg^*$.
Their variations $dF$, $dG$ are elements of $\gg$ defined as
$$
F(\bfS+{\bf\varepsilon})=F(\bfS)+\lan{\bf\varepsilon},dF(\bfS)\ran\,.
$$
Define the Poisson brackets as
$$
\{F(\bfS),G(\bfS)\}=\lan\bfS,[dF(\bfS),dG(\bfS)]\ran\,.
$$
The Jacobi identity for them follows from the Jacobi identity for $\gg$.
In particular, for the linear functions  $S_a=\lan\bfS,e_a\ran$ we find
$$
\{S_a,S_b\}=C_{ab}^cS_c\,.
$$
Thus, we come to the linear (Lie-Poisson) brackets on $\gg^*$.
These brackets are degenerated. The Casimir functions $C_s$ Poisson commute with functions
on $\gg^*$. The variety $C_s=const$ is a coadjoint orbit (\ref{co}). It
 is a symplectic manifold with the form
\beq{sfco}
\om=\lan \de(S_0g^{-1})\wedge \de g\ran\,.
\eq
\bigskip

Any $H\in C^\infty(\clR)$ defines a Hamiltonian vector field on $\clR\,$
$$
H\to\lan dH|\pi=\p_aH\pi^{ab}\p_b=\{H,\,\}\,.
$$
A Hamiltonian system is a triple $(\clR,\pi,H)$ with the Hamiltonian flow
$$
\p_tx^a=\{H,x^a\}=\p_bH\pi^{ba}\,.
$$

A Hamiltonian system is called {\it completely integrable}, if it satisfies the
following conditions
\begin{itemize}
  \item there exist $n$ Poisson commuting
Hamiltonians on $\clR$ ({\it integrals of motion})
$I_1,\ldots,I_n$
  \item Since the integrals commute the set

  \beq{lt}
  T_c=\{I_j=c_j\,,~(j=1,\ldots,n)\}
  \eq
   is invariant with respect to the Hamiltonian flows
  $\{I_j,~\}$. Then being restricted to $T_c\,$, $\,I_j(x)$
  should be functionally independent, i.e. $\det(\p_aI_b)(x)\neq 0$ for almost  all $x\in T_c$.
\end{itemize}

In this way we come to the hierarchy of commuting flows on $\clR$
\beq{hh}
\p_{t_j}\bfx=\{I_j(\bfx),\bfx\}\,,~~(j=1,\ldots,n)\,.
\eq

$T_c$ (\ref{lt}) is \emph{a Lagrangian} submanifold $T_c\subset\clR$, i.e. $\om$ vanishes on $T_c$.
If $T_c$ is compact and connected, then it is diffeomorphic to a $n$-dimensional torus.
Torus $T_c$ is called {\it the Liouville torus}.
Locally  there is a projection
 \beq{pro}
 p\,:\,\clR\to B\,,
 \eq
  where the Liouville tori are generic fibers and the base of fibration
  $B$ is parameterized by the values of the integrals.
 The coordinates on a Liouville torus ("the angle" variables) along with dual variables on $B$
("the action" variables) describe a linearized motion on the torus.
Globally, the picture can be more complicated. For some values of $c_j\,$ $\,T_c$ ceases to be a submanifold. In this way the action-angle variables are local coordinates.



Here we consider a complex analog of this picture (see, for example, \cite{Do}).
We assume that $\clR$ is a complex algebraic manifold and the symplectic
form $\om$ is a $(2,0)$ form, i.e. locally in the coordinates $(z^1,\bz^1,\dots,z^l,\bz^l)\,$
the form is represented as $\,\om=\om_{a,b}dz^a\wedge dz^b$. General fibers of
(\ref{pro})  are {\it abelian subvarieties} of $\clR$, i.e. they
are complex tori $\mC^l/\Lambda$, where the lattice $\Lambda$  satisfies the Riemann conditions \cite{GH}.
Integrable systems in this situation are called {\it algebraically
integrable systems}.


\subsection{Lax representation}

 The commonly accepted method
for constructing and  investigating integrable systems is
 based on {\it the Lax representation.} What is imported for us is that
 it reveals interrelations between
 classical integrable systems and gauge theories.

Let $\gg$ be a simple finite dimensional Lie algebra. The dual space $\gg^*$
can be identified with $\gg$ by means of a fixed Killing form $(~,~)$.
Introduce an additional parameter $z\in\mC$. It is called \emph{the   spectral parameter}.
The Lax matrix $L(\bfx,z)$ belongs to $\gg^*$ and depends on $z$
and on the dynamical variables $\bfx$.
Consider the integrable hierarchy (\ref{hh}).
 Assume that
the commuting flows (\ref{hh}) can be rewritten in the matrix form
\beq{Lax}
\p_{t_j}L(\bfx,z)=[L(\bfx,z),M_j(\bfx,z)]\,,
\eq
where $\,M_1(x,z),\ldots,M_n(x,z)$
be a set of $n$ matrices in the adjoint representation of $\gg$.

The system (\ref{Lax}) looks over-determined since
since we have infinite number of equations upon expanding  $L(\bfx,z)$ and $M_j(\bfx,z)$
in Laurent series in $z$. In fact, as it will be explained later,
  $L(\bfx,z)$ and $M_j(\bfx,z)$ are defined globally as meromorphic functions
  (more exactly sections of adjoint bundles) on a Riemann surface $\Si_g$
  of genus $g$.  The space of this sections is finite-dimensional and coincides with the phase  space $\clR$.

Let two integrable systems be described
by  two isomorphic sets of the action-angle variables.
In this case the integrable systems can be considered as equivalent. Establishing equivalence in
terms of angle-action variables is troublesome, but in terms of (\ref{Lax})
there it takes the form of the gauge equivalence.
Let $f$ be a non-degenerate $z$-dependent matrix.
The transformation
\beq{gau}
L'=f^{-1}Lf\,,~~M_j'=f^{-1}\p_{t_j}f+f^{-1}M_jf\,.
\eq
is called the gauge transformation because it preserves the Lax form (\ref{Lax}).
The flows (\ref{Lax}) can be considered as special gauge transformations
$$
L(t_1,\dots,t_l)=f^{-1}(t_1,\dots,t_l)L_0f(t_1,\dots,t_l)\,,
$$
where $L_0$ is independent on times and defines an initial data, and $M_j=f^{-1}\p_{t_j}f$.
Moreover, it follows from this representation that the quantities $\tr (L(\bfx,z))^j$
 are preserved by the flows and thereby can produce, in principle, Poison commuting integrals of motion. In what follows we will  construct $L$ in a such way that that $\tr (L(\bfx,z))^j$ being expanded in the basis of meromorphic functions on $\Si_g$
become commuting integrals.

 As we mentioned above,
it is  reasonable to consider two integrable systems to be equivalent if their Lax matrices
are related by a non-degenerate gauge transformation.

The gauge invariance of the Lax matrices allows one to define the spectral curve
\beq{spc}
\mathcal{C}=\{(\la\in\mC\,,z\in\Si)\,|\,\det(\la-L(x,z))=0\}\,.
\eq
The Jacobian of $\mathcal{C}$ \cite{GH} is an abelian variety of dimension $g$, where $g$ is the genus of $\mathcal{C}$.
If  $g=n=\oh\dim\,\clR$ then $\mathcal{J}$  plays the role of the Liouville torus
and the system is algebraically integrable. In generic cases $g>n$ and to prove algebraic
integrability one should find additional reductions of the Jacobians, leading to abelian spaces
of dimension $n$.


\section{Holomorphic bundles over Riemann surfaces}
\setcounter{equation}{0}

\subsection{Global description of holomorphic bundles}

Consider a Riemann surface $\Si_g$ of genus $g$.
Let  $\pi_1(\Si_g)$ be a fundamental group of $\Si_g$.
It has $2g$ generators $\{a_\al,b_\al\}\,$, corresponding to the
 fundamental cycles of $\Si_g$ with the relation
\beq{0.3}
\prod_{\al=1}^g [b_{\al},a_{\al}]=1\,,
\eq
where $[b_{\al},a_{\al}]= b_{\al}a_{\al}b_{\al}^{-1}a_{\al}^{-1}$ is
the group commutator.

Consider a finite-dimensional representation $\pi$ of a simple complex Lie group $G$
in a space $V$. Let $E_G$ is a principle $G$-bundle  over $\Si_g$. We
define a holomorphic $G$-bundle  $E=E_G\times_GV$ over $\Si_g$ using $\pi_1(\Si_g)$.
The bundle $E$ has the space of sections $\G(E)=\{s\}$,
where $s$ takes values in $V$.
Let $\rho$ be a representation of $\pi_1$ in $V$ such that $\rho(\pi_1)\subset\pi(G)$.
The bundle $E$ is defined by transition matrices of its sections
 around the fundamental cycles.
Let $z\in \Si_g$ be a fixed point. Then
\beq{sect}
 s(a_\al z)= \rho(a_\al)s(z)\,,~~
 s(b_\be z)= \rho(b_\be)s(z)\,.
\eq
Thus, the sections are defined by their quasi-periodicities on the fundamental cycles.
Due to (\ref{0.3}) we have
\beq{pi}
\prod_{\al=1}^g [\rho(b_{\al}),\rho(a_{\al})]=Id\,.
\eq

The $G$-bundles described in this way are topologically trivial. To consider
less trivial situation one should consider $G$-bundles where $G$ has a non-trivial
center. Centers of simple Lie groups are finite abelian groups (see Table 1. in Appendix). Let $\bar G$ be a complex simple simply-connected Lie group.
It can have a maximal nontrivial center. The centers are cyclic groups except Spin$_{4n}(\mC)$.
The factor group $G^{ad}=\bar G/\clZ(\bar{G})$ is the adjoint
group. The most known example is $\bG$=SL$(2,\mC)$. Its compact form
is SU$(2)$ and the center is $\mu_2=\mZ/2\mZ$ generated by the diagonal matrix
$\di(-1,-1)$. $G^{ad}=$SU$(2)/\mu_2=$SO$(3)$.

The  adjoint representations of $\bar G$ and $G^{ad}$ coincide.
In the cases $A_{n-1}\,$, $\,(n$ is a prime number), $B_n$, $C_n$,
$E_6$ and $E_7$ only  $\bar G$ and $G^{ad}$ there are only $\bar G$ and $G^{ad}$
two groups with the same Lie algebras.
In the rest cases there exist intermediate groups $G_{ad}\subset G\subset \bar G$,
since $\clZ(\bar{G}$ has a non-trivial subgroups and one can factorize $\bG$ with respect
to these subgroups.
Consider, for example, $\bar G=Spin_{4n}(\mC)$. It has a non-trivial center
\beq{cd2}
 \clZ(Spin_{4n})=(\mu^L_2\times\mu^R_2)\,,~~\mu_2=\mZ/2\mZ\,,
\eq
where three subgroups can be described in terms of their generators as
$$
\mu^L_2=\{(1,1)\,,(-1,1)\}\,,~~\mu^R_2=\{(1,1)\,,(1,-1)\}\,,~~\mu^{diag}_2=\{(1,1)\,,(-1,-1)\}\,.
$$
 Therefore there are three intermediate subgroups between $\bar G=Spin_{4n}(\mC)$
 and $G^{ad}$
\beq{hsog}
\begin{array}{ccccc}
   &   & Spin_{4n} &  \\
   & \swarrow & \downarrow & \searrow &  \\
  Spin_{4n}^{R}= Spin_{4n}/\G^L &   & SO(4n)= Spin_{4n}/\G^{diag}&  &Spin_{4n}^{L}=Spin_{4n}/\G^R\\
   & \searrow & \downarrow & \swarrow &  \\
   &  & G^{ad}=SO(4n)/(\mu^L_2\times\mu^R_2) &  &
\end{array}
\eq

Assume now that $G$ has a non-trivial center $\clZ(G)$.
Let $\zeta\in\clZ(G)$.
Replace (\ref{pi}) by
\beq{pi1}
\prod_{\al=1}^g [\rho(b_{\al}),\rho(a_{\al})]=\zeta\,.
\eq
Then the pairs $(\hat\rho(a_\al),\hat\rho(b_\be))$, satisfying (\ref{pi1}) cannot describe  transition matrices of $G$-bundle, but can serve as transition matrices of $G^{ad}=G/\clZ(G)$-bundle.
The bundle $E$ in this case is topologically non-trivial and
 $\zeta$ is called \emph{the characteristic class} $\bfc(E)$ of $E$.
 \footnote{Strictly speaking the characteristic class is element of the
 cohomology $H^2(\Si_g,\clZ(G))$. But in fact $H^2(\Si_g,\clZ(G))=\clZ(G)$.}

 For $Spin_{4n}$ bundles $\bfc(E)=\zeta$ in (\ref{pi1}) can be chosen as
 elements from three subgroups $\mu^L$, $\mu^R$, or
 $\mu^{diag}$. Then they give rise to three characteristic classes that
 are obstructions to lift  $Spin_{4n}^{R}$, $Spin_{4n}^{L}$ and $SO(4n)$ bundles to
 $Spin_{4n}$  (see (\ref{hsog})). The latter obstruction is related to the
 Stiefel-Whitney characteristic class \cite{EGH}.

The transition matrices can be deformed without breaking  (\ref{pi}) or
(\ref{pi1}).  Among these deformations are the gauge transformations
 $$
\rho(a_\al)\to f^{-1}\rho(a_\al)f\,,~~\rho(b_\be)\to f^{-1}\rho(b_\be)f\,.
$$
\emph{The moduli space of holomorphic bundles} $\clM_{g}$ are the space of deformations defined up to the gauge transformations.
 Its dimension is independent on the characteristic class and
is equal
\beq{mg}
\dim\,(\clM_{g})=(g-1)\dim\,(G)\,.
\eq
It means that the nonempty moduli spaces arise for the holomorphic bundles over
surfaces of genus $g>1$.

To include in the construction the surfaces with $g=0,1$
 consider a Riemann surface with $n$ marked points and attribute $E$ with a special structure at the  marked points. Let $B$ be a Borel subgroup of $G$.
We assume that the gauge transformation $f$ preserves the flag  variety $Fl=G/B$
(see Appendix). It means that
$f\in B$ at the marked points. It follows from (\ref{fld}) that
 \beq{mgn}
\dim\,(\clM_{g,n})=(g-1)\dim\,(G)+n\dim\, (Fl)=(g-1)\dim\,(G)+n\sum_{j=1}^l(d_j-1)\,.
\eq
In the important for applications case $g=1$, $n=1$
$\dim\,(\clM_{g,1})=\dim\,(Fl)$.

 Holomorphic sections can be described in another terms by a connection $d_{\bA}=\bp+\bA$. Assume that
 $\bA$ has the quasi-periodicities
 \beq{qpa}
 d_{\bA(a_\al z)}=\rho^{-1}(a_\al)d_{\bA(z)}\rho(a_\al) \,,~~~
d_{\bA(b_\be z)}=\rho^{-1}(b_\be)d_{\bA(z)}\rho(b_\be) \,.
\eq
The holomorphic sections are those that are annihilated  by $d_{\bA}$
\beq{hsec}
d_{\bA} s=0\,.
\eq


\subsection{Local description of holomorphic bundles and modification.}

There exists another description of a holomorphic bundles over $\Si_g$,
described, for example, in \cite{W1}.
Let $w_0$ be a fixed point on $\Si_g$
 and $D_{w_0}$ ($D^\times_{w_0}$) be a disc (punctured disc) with a center $w_0$
with a local coordinate $z$. Consider
a $G$-bundle $E=E_G\times_GV$ over  $\Si_g$.
It can be trivialized over  $D$ and over
$\Si_g\setminus w_0$. These two trivializations are related by a
$G$ transformation $\pi(g)$,
  holomorphic in $D^\times_{w_0}$, where $D_{w_0}$ and $\Si_g\setminus w_0$ overlap.
If we consider another trivialization over $D$
then $g$ is multiplied from the right by  $h\in G$.
Likewise, a trivialization over $\Si_g\setminus w_0$ is determined up to the
multiplication on the left $g\to hg$ , where $h\in G$ is holomorphic on
 $\Si_g\setminus w_0$.
Thus, the set of isomorphism classes
$G$-bundles are described as a double-coset
\beq{mshb}
G(\Si_g\setminus w_0)\setminus G(D^\times_{w_0}) /G(D_{w_0})\,,
\eq
where $G(U)$ denotes the group of $G$-valued holomorphic functions
on $U$.

To  define a $G$-bundle over $\Si_g$ the transition matrix $g$ should have
 a trivial monodromy around $w_0$ $g(ze^{2\pi i})=g(z)$ on the punctured disc $D^\times_{w_0}$.
 But if the monodromy is nontrivial
 $$
 g(ze^{2\pi i})=\zeta g(z)\,,~~\zeta\in\clZ(G)\,,
 $$
 then $g(z)$ is not a transition matrix. But it can be considered as
  a transition matrix for the $G^{ad}$-bundle, since in the group $G^{ad}$
  the center does not act.
  This relation is similar to (\ref{pi1}).

  Our aim is to construct a new bundle $\ti E$ with a non-trivial characteristic from $E$.  This procedure is called \emph{a modification} of bundle $E$.
Smooth gauge transformations cannot change a topological type of bundles.
The modification is defined by a singular gauge transformation at some point,
say $w_0$. Since it is a local transformation we replace $\Si_g$ by a sphere
$\Si_0=\mC P^1$, where $w_0$ corresponds to the point $z=0$ on $\mC P^1$. Since $z$ is local coordinate, we
can replace
 $G(\Si_g\setminus w_0)$ in (\ref{mshb}) by the group $G(\mC((z)))$. It is the
group of Laurent series with $G$ valued coefficients. Similarly, $G(D_{w_0})$ is replaced
on $G(\mC[[z]])$. It is a power series. It is clear from this description of the
moduli space of bundles over $\mC P^1$ that it is a finite dimensional space.

Transform $g(z)$ by multiplication from the right on $h(z)$
$g(z)\to g(z)h(z)$ that singular at $z=0$.
It is the singular gauge transformation mentioned above. Due to definition of
$g(z)$, $\,h(z)$ is defined up to the multiplication from the right on $f(z)\in G(\mC[[z]])$.
On the other hand, since $g(z)$ is defined up to the multiplication from the right on
element $G(\mC[[z]])$, $h(z)$ is element of the double coset
$$
G(\mC[[z]])\setminus G(\mC((z)))/G(\mC[[z]])\,.
$$
In particular $h(z)$ is defined up to conjugation. It means that as a representative
of this double coset one can take
 a co-character (\ref{cocharc}) $h(z)\in t(G)$.
\beq{mtm}
g(z)\to g(z)z^{\ga}\,,~~~(z^{\ga}=\bfe(\ln\,(z)\ga))\,,
\eq
where $\ga$ belongs to the coweight lattice $(\ga=(m_1,m_2,\ldots,m_l)\in P^\vee)$
(\ref{cwl}).
The monodromy of $z^{\ga}$ is an element of $\clZ(\bG)$ (\ref{center}). In this way we come to a new bundle $\ti E$ with a non-trivial characteristic class.
The bundle $\ti E$ is called \emph{the modified bundle}.
It is defined by the new transition matrix (\ref{mtm}).
If $\ga\in Q^\vee\,$ then $\zeta=1$ and the modified bundle is trivial.

This transformation of the bundle $E$ corresponds to transformations of its sections
$\ti E$
\beq{mod1}
\G(E)\,\mapright{\Xi(\ga)} \,\G(\ti E)\,,~~
(\Xi(\ga)\sim\pi(z^{m_1},z^{m_2}\ldots z^{m_l}))\,.
\eq
We call this modification  of type $\ga=(m_1,m_2,\ldots,m_l)$.
Another name of the modification is \emph{the Hecke transformation}.
Usually the algebra of functions on double coset of type $K\setminus G/K$ is called
a Hecke algebra.
The case $G = {\rm SL}(2,\mathbb{Q})$ and $K ={\rm SL}(2,\mZ)$
leads to the abstract ring of Hecke operators in the theory of modular forms, which gave the name to Hecke algebras in general.
In field-theoretical terms the Hecke transformation  corresponds to \emph{the t'Hooft operator,} generating by monopoles (see below).

Consider the action of modification on sections (\ref{mod1})  in more details.
Choose a Cartan subalgebra $\gH$ in $\gg$ and the corresponding weight basis
$(|\nu_1\ran,\ldots,|\nu_M\ran)$ in $V\,$  $\,(M=\dim\,(V))$. It means that for $\bfx\in\gH\,$
$\,\pi(\bfx)|\nu_j\ran=\lan\bfx,\nu_j\ran|\nu_j\ran$.
The weights belong to the weight diagram defined by the highest weight $\nu\in P$   of $\pi$
\beq{nu}
\nu_j=\nu-\sum_{\al_m\in\Pi}c_j^m\al_m\,,~~~c_j^k\in\mZ\,,~c_j^k\geq 0\,.
\eq
Let us choose a trivialization of $E$ over $D$ by fixing this basis.
Thereby, the bundle $E$ over $D$ is represented by a sum
of $N$ line bundles $\clL_1\oplus\clL_2\oplus\ldots\oplus\clL_M$.
Cartan subgroup $\clH$ acts in this basis in a diagonal way: for
$\bfs=(|\nu_1\ran,\ldots,|\nu_M\ran)$
$$
\pi(h)\,:\,|\nu_j\ran\to \bfe\lan\bfx,\nu_j\ran |\nu_j\ran\,,~~~h=\bfe\,(\bfx)\,,~~\bfx\in\gH\,,
~~~(\bfe(x)=\exp\,(2\pi i\bfx))\,.
$$
Assume for simplicity that in (\ref{mtm}) $g(z)=1$. Then
 the modification transformation (\ref{mod1}) of the sections assumes the form
 \beq{mod}
\Xi(\ga)\,:\, |\nu_j\ran\to z^{\lan \ga,\nu_j\ran} |\nu_j\ran\,,~~j=1,\ldots,M\,.
\eq
It means that away from the point $z=0$ where the transformations are singular sections of $\ti E$ is the same as $E$. But near $z=0$ they are singular with the leading terms
$|\nu_j\ran\sim c_jz^{-\lan \ga,\nu_j\ran}$.

It sufficient to consider the case when  $\ga=\varpi^\vee_i$ is a fundamental coweight and $\pi$ is
a fundamental representation $\nu=\varpi_k$.
Then from (\ref{nu}) we have
$$
z^{\lan \ga,\nu_j\ran}=z^{\lan\varpi^\vee_i,\varpi_k-\sum_{\al_m\in\Pi}c_j^m\al_m \ran}\,.
$$
The weight $\varpi_k$ can be expanded in the basis of simple roots  $\varpi_k=\sum_kA_{km}\al_m$,
where $ A_{jk}$ is the inverse Cartan matrix ($A_{jk}a_{ki}=\de_{ji}$).
Its matrix elements are rational numbers with the denominator $N=ord\,(\clZ)$.
Then from (\ref{cwl})
$$
z^{\lan \ga,\nu_j\ran}=z^{A_{ik}-c_j^i}\,,~~c_j^i\in\mZ\,.
$$
It implies that  the modification can produce a non-trivial branching of sections
$z^{\lan \ga,\nu_j\ran}\sim z^{A_{ik}}$ if $A_{ik}$ is non-integer. Note, that the branching does not happen for $G^{ad}$-bundles, because the corresponding  weights $\nu_j$ belong to the root lattice $Q$ and thereby $\lan \ga,\nu_j\ran\in\mZ$.

It is possible
to go around the branching by multiplying the sections on a scalar matrix of the form
$\di(z^{-A_{ik}},\ldots,z^{-A_{ik}})$. This matrix  no longer belongs to the representation of
$\bG$, because it has the determinant $z^{-MA_{ik}}$ $(M=\dim\,V)$. It can be checked that in  particular cases that $MA_{ik}$ is an integer number, though I don't know how to prove it in general case.

If $G=\SLN$ the scalar matrix  belongs to $\GLN$. Thereby, after this transformation we come to a $\GLN$-bundle.
But this bundle is topologically non-trivial, because it has a non-trivial degree of its determinant bundle. In this way the characteristic classes for the $\SLN$-bundles
are related to another topological characteristic, namely to  degrees of the  $\GLN$-bundles.

It is possible to construct
an analog of $\GLN$ for other simple groups \cite{LOSZ}. We call them the \emph{conformal
versions} of simple groups, since they can be described as groups preserving some
forms up to dilatations, likewise $\GLN$ preserve the volume form in $\mC^N$ up
to multiplications. In  \cite{LOSZ} we describe also interrelations between the
characteristic classes for simple groups and degrees of the related determinant bundles
of their conformal versions.


\section{Higgs bundles and integrable systems.}
\setcounter{equation}{0}

 The Higgs field $\Phi$ is an element of
$\Om^{0}(\Si_g,ad(E)\otimes K_\Si)$, where $K_\Si$ is the canonical line bundle of $\Si_g$. It means that
locally $\Phi$ is represented in the form $\Phi(z,\bz)dz$, where $\Phi$ takes value in the adjoint representation of $\gg$. Assume that $\Phi$ satisfies (\ref{qpa}).
The pair
\beq{shs}
\clR^{H}=(\Phi,d_{\bA})
\eq
is called \emph{the Higgs $G$-bundle} over  $\Si_{g}$.

The fields
 $(\Phi,d_{\bA})$ are  coordinates  in the infinite dimensional
 cotangent bundle $T^*\{E\}$ to the space of holomorphic bundles $\{E\}$
 defined by $d_{\bA}$. The Higgs field plays the role of the cotangent vector.

The Killing form $(~,~)$ in $\gg$ equips
  $\clR^{H}$ with the symplectic form
\beq{sfh}
\Om=\int_{\Si_g}(\de\Phi\wedge\de\bA)\,.
\eq
It  is a canonical symplectic form on the  cotangent bundle  $\clR^{H}$.

 Introduce a complex structure in the infinite-dimensional space
 $\clR^{H}$. The fields $\Phi$ and $\bA$ are holomorphic coordinates in this structure.
 Then $\Om$ is the $(2,0)$-form in this structure, what we need for the algebraic
 integrability.

The form  (\ref{sfh}) is invariant with respect to the
gauge transformations $\clG$
\beq{gth}
\Phi\to \Phi^f= f^{-1}\Phi f\,,~~~\bA\to \bA^f=f^{-1}\bA f+f^{-1}\bA f\,.
\eq
Their infinitesimal form
$$
\de_\epsilon\bA=\bp\epsilon+[\bA,\epsilon]\,,~~
\de_\epsilon\Phi=[\Phi,\epsilon]\,,~~\epsilon\in Lie(\clG)\,.
$$
is generating by an analog of the  Gauss law.
Namely, it is easy to check that
$$
\de_\epsilon\bA=\{\Upsilon,\bA\}\,,~~~
\de_\epsilon\Phi=\{\Upsilon,\Phi\}\,,~~~~
\Upsilon=\int_{\Si_g}(\epsilon,d_{\bA}\Phi)\,,
$$
where $\{~,~\}$ is the Darboux brackets in $\clR^H$, inverse to $\Om$.
Putting
\beq{glh}
d_{\bA}\Phi=0
\eq
we impose the first class constraints.
In a general setting the Gauss law is called \emph{the moment constraints}.
The operator $d_{\bA}$ defines a complex
structure on $E$ and (\ref{glh}) means that $\Phi$ is a holomorphic section of $ad(E)\otimes K_\Si$ (\ref{hsec}).

Along with a gauge fixing constraints, the moment constraints form the second class constraints.
Physical degrees of freedom (the moduli space of the Higgs bundles) are defined as
$$
{\cal R}^H(\Si_g)/({\rm Gauss~ law})+({\rm gauge~fixing})=(d_{\bA}\Phi=0)/\clG={\cal R}^H(\Si_g)//\clG\,.
$$
The quotient space with respect to these constraints is called \emph{the symplectic quotient}. We call it
the moduli space of Higgs bundles
\beq{redu}
T^*{\cal M}_g={\cal R}^H(\Si_g)//\clG\,.
\eq
This symplectic quotient space is finite-dimensional
$$
\dim\,(T^*{\cal M}_g )=2(g-1)\dim\,(G)\,.
$$
Its dimension is twice of dimension of $\clM_g$ (\ref{mg}).


\subsection{Higgs bundles on Riemann surfaces with marked points}

To deal with a sphere $g=0$ and a torus $g=1$ consider Riemann surfaces with
marked points. Let $\Si_{g,n}$ be a Riemann surface of genus $g$ with $n$ marked points. Attribute  coadjoint orbits $\clO_a$ to the
 marked point $z_a$ $(a=1,\ldots n)$. The space
\beq{hps}
\clR^H=\{(\bA\,,\Phi\,,\clO_a\,,~a=1,\ldots n)
\eq
is symplectic with the form
$$
\Om+\sum_{a=1}^n\om_a\,,
$$
 where $\om_a$ is the symplectic form on $\clO_a$ (\ref{sfco}).
 It is possible to introduce local holomorphic coordinates on the
 coadjoint orbits.
 We do not need in their description.

 The gauge transformations (\ref{gth}) should be completed by
 $$
 \bfS_a\to  \bfS_a^f=f^{-1}(z_a,\bz_a)\bfS_af(z_a,\bz_a)\,.
 $$
It can be find that the moment constraints in this case take the form
\beq{mcmp}
d_{\bA}\Phi=\sum_a\bfS_a\de(z_a)\,,~~~\bfS_a\in\clO_a\,,
\eq
where $de(z_a)$ is the delta-function with support at $z_a$.
In other words $\Phi$ is a meromorphic section of $End E\otimes K$ with simple
poles
$$
Res\,\Phi|_{z=z_a}=\bfS_a\,.
$$
Dimension of the symplectic quotient in this case is
\beq{drs}
\dim\,(T^*{\cal M}_{g,n})=2(g-1)\dim\,(G)+\sum_{a=1}^n\dim\,(\clO_a)=
(2(g-1)+n)\dim\,(G)-nl\,,
\eq
where $l=rank\,G$.


\subsection{Hitchin integrable systems}

Since $T^*{\cal M}_{g,n}$ is a result of symplectic reduction it
is a symplectic manifold. The corresponding  Poisson brackets are Dirac
brackets \cite{Di} obtained from the canonical brackets on $\clR^{H}$.
To construct an integrable hierarchy one should find $N=\oh\dim\,(T^*{\cal M}_{g,n})$
independent Poisson commuting Hamiltonians. They are constructed from
 invariant polynomials on $\gg$ making from the Higgs field \cite{Hi1}.
 The invariant polynomials  have the form  $\tr(\Phi^{d_j})\,(j=1,\ldots,l)$ where $d_j$ are degree of invariants  of $\gg$.
  For $D_{2k}$ there are two invariants of order $k$. In addition to $\tr( \Phi^{k})\,$
  it is the pfaffian of $\Phi$. The polynomials are
   $(d_j,0)$-forms $(\Phi^{d_j})\in\Omega^{(d_j,0)}(\Si_{g,n})$
with holomorphic poles of order $d_j$ and less at the marked points.
To construct Hamiltonians  one should integrate
them over $\Si_{g,n}$.  For this purpose one should prepare $(1,1)$-forms from the $(d_j,0)$-forms. It can be done by  smooth
$(1-d_j,1)$-differentials  vanishing at the marked points.
 Locally, they are represented as
 $\mu_j=\mu_j(z,\bz)\left(\frac{\p}{\p z}\right)^{j-1}\otimes d\bz$.
 In other words $\mu_j$ are $(0,1)$-forms taking values in $j$ degree $\clT^{\otimes j}$
 of vector fields on $\Si_{g,n}$.
For example, $\mu_2$ is the Beltrami differential.
The product $ (\Phi^j)\mu_j$ can be integrated over the surface.

The space  $\Omega^{(1-d_j,1)}(\Si_{g,n})$ of $(1-d_j,1)$-differentials
is infinite-dimensional. We identify its elements by diffeomorphisms of $\Si_{g,n}$.
 To construct Hamiltonians one can take elements of the quotient of $\Omega^{(1-d_j,1)}(\Si_{g,n})$
 with respect to this action. It is a finite-dimensional space with a local description as
 the cohomology space $H^1(\Si_{g,n}, \mathcal{T}^{\otimes d_j-1})$ of
$\Si_{g,n}$ with coefficients in tensor degrees of vector fields $\clT$
 on $\Si_{g,n}$.
  This space has dimension
\beq{dime}
n_j=\dim H^1(\Si_{g,n},\clT^{\otimes(j-1)})=
  (2d_j-1)(g-1)+(d_j-1)n\,.
\eq

Let  $\mu_{j,k}$ be a basis in $H^1(\Si_{g,n},\clT^{\otimes(j-1)})\,$, $~(k=1,\ldots,n_j)$.
The integrals
\beq{int}
I_{j,k}=\f1{d_j}\int_{\Si_{g,n}}
\mu_{j,k}(\Phi^{d_j})\,,~~j=1,\ldots l\,.
\eq
define Poisson commuting Hamiltonians.

It follows from (\ref{dime}), (\ref{dig}) and (\ref{dio}) that the number of integrals is equal to
$$
\sum_{j=1}^ln_j=(g-1)\sum_{j=1}^l(2d_j-1)+n\sum_{j=1}^l(d_j-1)=
(g-1)\dim\,G+\oh n\dim\clO\,.
$$
It is the dimension of the moduli space of holomorphic bundles ${\cal M}_{g,n}$
(\ref{mgn}). It is equal to
the half of $\dim\,({\cal M}_H(\Si_{g,n}))$ (\ref{drs}) for generic orbits at all
marked points.

The Hamiltonians $I_{j,k}$ define a free motion on $\clR^H$
\beq{uem}
\p_{j,k}\Phi=\{I_{j,k},\Phi\}=0\,,~~\p_{j,k}\bA=\{I_{j,k},\bA\}=\Phi^{j-1}\mu_{j,k}\,.
\eq
These equations being trivial on $\clR^H$ become meaningful on $T^*{\cal M}_{g,n}$
and define integrable hierarchies.

Since $d_{\bA^f}\Phi^f=f^{-1}(d_{\bA}\Phi)f$ and $f$ vanishes at the marked points,
to describe   $T^*{\cal M}_{g,n}$ it is possible  first to fix a gauge and then to solve
the moment constraint equation. Let us
 choose a gauge $\bA_0=f^{-1}\bA f+f^{-1}\bA f$. Then by the same gauge transformation define the transformed Higgs field
\footnote{Generically, $f$ is defined up to residual gauge transformations.
 We will not stop here on this point.}
\beq{lh}
L=f^{-1}\Phi f\,.
\eq
It follows from the first equation (\ref{uem}) that $L$ is the Lax operator
\beq{leq}
\p_{j,k}L=[L,M_{j,k}]\,,~~~M_{j,k}=f^{-1}\p_{j,k}f\,.
\eq
The form of the M-operator can be extracted from the second equation in  (\ref{uem}).
In terms of the Lax operator (\ref{mcmp}) assumes the form
\beq{mcmp1}
d_{\bA_0}L=\sum_a\bfS_a\de(z_a)\,.
\eq
Here we keep the notation $\bfS_a$ for the gauge transformed elements.
Then the moduli space of the Higgs bundles ${\cal M}_H(\Si_g)$ coincides with the space of solutions of this equation.
The Lax operator is a meromorphic section of adjoint bundle, specified by its monodromies (\ref{sect}). They can satisfy (\ref{pi}) or (\ref{pi1}).
Therefore $L$ depends on
the moduli of bundles and on residues $\bfS_a$ of its poles. In a concrete interpretation $\bfS_a$ are identified with spin variables, while local coordinates on
the moduli space $\clM_{g,n}$ play the role of coordinates of particles. Their momenta are moduli of solutions of (\ref{mcmp1}).

 It can be proved that the Hamiltonians
(\ref{int}) being restricted on $T^*{\cal M}_{g,n}$
$$
I_{j,k}=\f1{d_j}\int_{\Si_{g,n}}
\mu_{j,k}(L^{d_j})
$$
continue to Poisson commute. Thus,  we come to completely integrable systems.
The set $B=\{I_{j,k}\}$ forms the base of the fibration (\ref{pro}).
To prove algebraic integrability one should construct the Liouville fibres
related to the spectral curve
$$
\clC(\la,z)=0\,,~~~\clC(\la,z)=\det(\la-L(z))\,.
$$
 For curves without marked points it was done
in \cite{Hi1,Fa}.


\subsection{Symplectic Hecke Correspondence}

Consider two Higgs bundles $(E,\Phi)$ and $(\ti E,\ti\Phi)$, where $\ti E$ is the
$\Xi$ modification of $E$ of type $\ga$ (\ref{mod}).
 The modification acts on the Higgs bundles as
$$
(E,\Phi)\stackrel{\Xi}{\rightarrow}
(\tilde{E},\ti{\Phi})\,,
$$
\beq{him}
\Xi\Phi=\ti{\Phi}\Xi\,,~~\Xi\ti{\bA}=\bp\Xi+\bA\Xi\,.
\eq
We call this transformation \emph{the Symplectic Hecke Correspondence}
of the Higgs bundles,
since it is a symplectomorphism of two Higgs bundles with symplectic forms (\ref{sfh}).

The Higgs fields $\Phi$ and $\ti{\Phi}$ should be holomorphic
with prescribed simple poles at the marked points.
The holomorphity of the Higgs field put restrictions on its form.
Decompose $\Phi$ and $\ti\Phi$ in the Chevalley basis (\ref{CD}), (\ref{CBA})
$$
\Phi(z)=\Phi_{\gH}(z)+\sum_{\al\in R}\Phi_\al(z)E_\al\,,~~~
\ti\Phi(z)=\ti\Phi_{\gH}(z)+\sum_{\al\in R}\ti\Phi_\al(z)E_\al\,.
$$
Expand $\al$ in the basis of simple roots (\ref{rde})
 $\al=\sum_{j=1}^ln_j^\al\al_j$ and $\ga$
in the basis of fundamental coweights $\ga=\sum_{j=1}^lm_j\varpi_j^\vee$, such that
$\lan\ga,\al_j\ran\geq 0$ for simple $\al_j$.
\footnote{Such weights are called the dominant weights.}
  Then
$\lan\ga,\al\ran=\sum_{j=1}^lm_jn_j^\al$ is an integer number.
From (\ref{him}) and (\ref{crsr})  we find
\beq{shc}
\ti\Phi_{\gH}(z)=\Phi_{\gH}(z)\,,~~~
\ti\Phi_\al(z)=z^{\lan\ga,\al\ran}\Phi_\al(z)\,.
\eq
 In a neighborhood of $z=0$ $\Phi(z)$ should have the form
$$
\Phi_\al(z)=a_{\lan\ga,\al\ran}z^{-\lan\ga,\al\ran}+
a_{\lan\ga,\al\ran+1}z^{-\lan\ga,\al\ran+1}+
\ldots\,,~~~(\al\in R^-)\,,
$$
otherwise the transformed field becomes singular.
 It means that a modification of a Higgs bundle in a point $z_0$ of $\Si$
is not arbitrary, but depends on the local behavior of the Higgs
field (the Lax operator).
Thus, there are $\sum_{\al\in R^+}\lan\ga,\al\ran$ constraints on the Higgs field.


\subsection{Examples}

For elliptic curves $g=1$ solutions of (\ref{pi1}) can be found explicitly \cite{LOSZ}.
We identify $\Si_1$ with the quotient $\mC/(\tau\mZ+\mZ)$, where $\tau$ is the
modular parameter ($Im\,\tau>0$).
It allowed us to define the Lax matrices and in this way the Hamiltonians for any
Lie algebra and an arbitrary characteristic class.
Consider, for simplicity, an elliptic curve with one marked point. If the
 characteristic class is trivial, then the corresponding integrable system is well
 known. It is the elliptic Calogero-Moser system with spin. The system describe
 $l=$rank$(G)$ particles and $\dim\,(\clO)$ "spin" variables subjected to $2l$ constraints
 The phase space of the system is
 $$
 (\bfv,\bfu)\in T^*\gH\,, ~~\ti\bfS\in\clO//\clH\,,
 $$
where $\bfu=(u_1,u_2,\ldots,u_l)$ are coordinates of particles,
$\bfv=(v_1,v_2,\ldots,v_l)$ - their momenta.
As for the symplectic quotient $\clO//\clH$ it is described by the local coordinates
 $\bfS\in \gg^*$, subject to the moment constraints
 $\ti\bfS|_{\gH}=0$ with respect to the conjugation of $\clO$ by the Cartan subgroup $\clH$, and the gauge fixing constraints $\ti S_\al=1$ for simple roots $\al\in\Pi$.

The phase space has the following interpretation in terms of Higgs bundles.
The spin variables $\ti\bfS$ are residue of the Higgs field satisfying the just mentioned constraints.
The coordinate variables $\bfu$ describe the moduli space of the bundle, while their momenta parameterize solutions of the moment constraint equation (\ref{mcmp1}). Note that for
the  trivial characteristic class $\bfu\in\gH$.

The Hamiltonian takes the form
$$
H^{CM}=\oh(\bfv,\bfv)+\sum_{\al\in R^+}\ti S_\al\ti S_{-\al}\wp(\bfu,\al)\,.
$$
Here $\wp(x)$ is the Weierschtrass function. It is a double-periodic meromorphic function in
the fundamental domain $(1,\tau)$.

For nontrivial characteristic classes the phase space has the same dimension, since
we the corresponding systems are symplectomorphic, but it has a different structure.
The phase space is described in \cite{LOSZ}.
There exists a correspondence between an element of the center $\zeta$
 of $G$ and a Cartan subgroup $\clH_{\zeta}\subset\clH$.
 It is described in the following way. There are two fundamental cycles
 for $g=1$ $a\sim 1$, $b\sim\tau$.
 The solutions of (\ref{pi1}) can be taken in the
 form $\rho_1\in\clH$ and $\rho_\tau$ as a special Weyl transform, that defines a symmetry of the extended Dynkin diagram. Then the subgroup $\clH_{\zeta}$ is the invariant Cartan subgroup $\rho_\tau h\rho_\tau ^{-1}=h$ for
$h\in\clH_{\zeta}$.
 The moduli space of holomorphic bundles over $\Si_\tau$ with characteristic class
 $\zeta$
 are elements of the Cartan Lie subalgebra $\gH_\zeta=Lie(\clH_{\zeta})$.
The configuration space of particles is a subset $C$ of $\gH_{\zeta}$. Then the phase space takes the following form
$$
(\bfv,\bfu)\in T^*C\,, ~~\ti\bfS\in\clO//\clH_\zeta\,.
$$
Let $\ti R=R\cap\gH_\zeta\subset R$ be a subsystem of roots
and $H^{CM}_{\ti R}$ be the corresponding Calogero-Moser Hamiltonian.
We call  the subgroup $\ti G\subset G$ generated by  ${\ti R}$ the unbroken subgroup,
because the
whole Hamiltonian $H_\zeta$ has the form
$$
H_\zeta=H^{CM}_{\ti R}+\ti H\,,
$$
where $\ti H$ has the form of interacting tops. It is too long to write it here
explicitly. Detail can be found in   \cite{LOSZ}.

For example, for $G=\SLN$ and $\zeta=\di(\exp\,2\pi i\frac{N-1}N,\exp\,-2\pi i\f1{N},
\ldots,\exp\,-2\pi i\f1{N})$ generates the center $\mu_N$ of $\SLN$.
The transition matrices, satisfying (\ref{pi1}) are the t'Hooft matrices
$$
\rho_1=\di(1\,,\om\,,\ldots,\om^{N-1})\,,~~(\rho_\tau)_{jk}=\de_{j,k-1}\,,~(mod\,N)\,,
~~\om=\exp\,2\pi i\f1{N}\,,.
$$
It easy to find that the moduli space is empty in this case.
\footnote{This happens only for $G=\SLN$. For other groups the moduli spaces are non-empty. For this reason the so-called non-dynamical $R$-matrices exist only for $\SLN$.}
 $H_\zeta=\ti H$ is the Hamiltonian of the so-called elliptic top.
 For $N=2$ it is the standard Euler top.
 As it was explained above, the systems with different characteristic classes are
 symplectomorphic.
The symplectic Hecke correspondence between the $\SLN$ elliptic Calogero-Moser
system and the elliptic top was constructed in \cite{LOZ2}.

Another example is the group $E_6$ with the center $\mu_3$. In this case $\ti R=R_{G_2}$.
The unbroken subgroup is $G_2$. Thus, instead of six interacting $E_6$ particles only
two survive and we come to the Hamiltonian
$$
H_\zeta=H^{CM}_{G_2}+\ti H\,.
$$
Again the  $E_6$ elliptic Calogero-Moser system and the last system are symplectomorphic.

For the group $Spin_{4n}$ we obtained four types of integrable systems related to
to the scheme (\ref{hsog}).


\section{Monodromy preserving equations}
\setcounter{equation}{0}

There exist a wide class of  Hamiltonian systems related to integrable systems.
They  are described in terms of the Isomonodromy problem defined for a linear
matrix equations on $\Si_{g,n}$.  The corresponding
nonlinear equations are the monodromy preserving conditions. They
are equations of motion for non-autonomous Hamiltonian systems. We will see that
they share some common features with the Hitchin integrable systems.
The most familiar
examples are the Painlev\'{e} equations and the Schlesinger system.
It turns out that integrable systems is a sort of quasi-classical limit of
these  non-autonomous Hamiltonian systems (Section 5.3).

\subsection{Flat bundles and deformed complex structures}

Let $(z,\bz)$ be local coordinates on  $\Si_{g,n}$ and
$\p_{z}+\clA_z$, $\partial_{\bz}+\clA_{\bz}$  are corresponding component of
connection in the introduced before  bundle $E$. Consider the space of connections
completed as before by coadjoint orbits at the marked points
\beq{pfs}
\clR=\{\clA_z\,,\,\clA_{\bz}\,,\,\clO_a\,,~a=1,\ldots,n\}\,.
\eq
It can be equipped with the symplectic form
\beq{omega}
\Om=\Om^{\clA}+\sum_{a=1}^n\om_a\,,~~~
\Om^{\clA}=\int_{\Si_g}(\de(\clA_z)\wedge\de(\clA_{\bz}))\,.
\eq
The form $\Om^{\clA}$ is independent on the complex structure $(z,\bz)$ and can be
 rewritten as
\beq{cla}
\Om^{\clA}=\int_{\Si_g}(\de(\clA)\wedge\de(\clA))\,.
\eq
To describe a nontrivial dynamical systems deform $\p_{\bz}$ by mixing it with $\p_z$
 as $\p_{\bz}+\mu\p_z$,
 where $\mu\in\Om^{(-1,1)}(\Si_{g,n})$ is the defined above Beltrami differential on $\Si_{g,n}$.
The Beltrami differential can be defined in terms of local substitution of coordinates
$$
w=z-\epsilon(z,\bz)\,,~~\bar w=\bz-\overline{\epsilon(z,\bz)}\,.
$$
Then for small $\ep\,$
$\,\mu=\p_{\bz}\ep$.
We consider the Beltrami differential to be  equivalent to zero if $\ep$ is continued to a global  diffeomorphism  of $\Si_{g,n}$.
The equivalence relations in $\Om^{(-1,1)}(\Si_{g,n})$ is the
moduli space of complex structures on $\Si_{g,n}$. The tangent space to the
moduli space is the Teichm\"{u}ller space ${\cal T}_{g,n}\sim
H^1(\Si_{g,n},T\Si_{g,n})$.
We have found the dimension of this space (see (\ref{dime}) for $d_j=2$)
\beq{2.6}
d=\dim {\cal T}_{g,n}=3(g-1)+n.
\eq

If $(\mu_1^0,\ldots,\mu_l^0)$ is the basis in the vector space
$H^1(\Si_{g,n},T\Si_{g,n})$, then
\beq{2.7}
\mu=\sum_{s=1}^dt_s\mu_s^0\,.
\eq

In this construction it is convenient to deform $\p_{\bz}$ and do not touch $\p_z$. To this end we use non-holomorphic substitution
$$
w=z-\ep(z,\bz),~~~
\bar{w}=\bz\,.
$$
Then we come to connections
$$
\p_z+\clA_z\,,~~~ \p_{\bz}+\mu\p_z+\clA_{\bz}
$$
and $\Om^{\clA}$ in (\ref{omega}) is replaced on
$$
\Om^{\clA}=\int_{\Si_g}(\de(\clA_z)\wedge\de(\clA_{\bz}-\mu \clA_z))=
\int_{\Si_g}(\de(\clA_z)\wedge\de(\clA_{\bz}))-(\clA_z\de(\clA_z))\wedge\de\mu\,.
$$
 The second term is interpreted as
$\de H\wedge\de t$, where the Hamiltonian $H$ is defined below (\ref{hs}). Thus, $\mu$ corresponds to time variables.
Taking into account (\ref{2.7}) rewrite the symplectic form on $\clR$ as
the Cartan-Poincar\'{e} invariant \cite{Ar1}
\beq{esf}
\Om=\int_{\Si_g}(\de(\clA_z)\wedge\de(\clA_{\bz}))
+\sum_{a=1}^n\om_a-(\clA_z\de(\clA_z))\wedge \sum_{s=1}^l\de t_s\mu_s^0=
\Om_0-\sum_{s=1}^l\de H_s\wedge\de t_s\,,
\eq
where
\beq{hs}
\Om_0=\int_{\Si_{g,n}}(\de(\clA_z)\wedge\de(\clA_{\bz}))+\sum_{a=1}^n\om_a\,,
~~~ H_s=\oh\int_{\Si_{g,n}}(\clA_z,\clA_z)\mu_s^0\,.
\eq
The form  is defined on the bundle $\clP\to{\cal T}_{g,n} $ over
the  Teichm\"{u}ller space  ${\cal T}_{g,n}$ with fibers $\clR$ (\ref{pfs}).
The form $\Om$ is degenerated on $d=\dim {\cal T}_{g,n}$ vector fields $D_s$:~ ($\Om(D_s,\cdot)=0$)
$$
D_s=\p_{s}+\int_{\Si_{g,n}}(\clA_z\frac{\de}{\de\clA_{\bz}})\mu^0_s=
\p_{s}+\{H_s,~\}_{\Om_0}\,,~~~(\p_s=\p_{t_s})\,.
$$
Here the Poisson brackets are inverse to the non-degenerate form
$\Om_0$ on the fibers.
 The vector fields $D_s$ define the
equations of motion for any function $f(\clA_z,\clA_{\bz},\bfS_a,\mu)$ on ${\cal P}$
$$
\frac{df}{dt_s}=\p_{t_s}f +\{H_s,f\}_{\Om_0}\,.
$$
In particular,
\beq{2.14}
\p_s\clA_z=0\,,~~~\p_s\clA_{\bz}=\clA_{z}\mu^0_s\,, ~~~\p_s\bfS_a=0\,,
\eq
and therefore again we come to a free motion
$$
\clA_z(\mu)=\clA_z^0,~~\clA_{\bz}(\mu)=\clA_{\bz}^0+\mu \clA_z^0\,.
$$

In addition, there are the consistency conditions for the Hamiltonians
\beq{WE}
\p_sH_r-\p_rH_s+\{H_r,H_s\}_{\Om_0}=0\,.
\eq

Let $\Psi$ be a fundamental solution of the linear system ($\Psi\in \Om^{(0)}(\Si_{g,n},{\rm Aut} E)$)
\beqn{2.15}
\left\{
\begin{array}{ll}
1. &  [\p_z+\clA_z,\Psi]=0\,, \\

2. &  [\p_{\bz}+\sum_{s=1}^lt_s\mu^0_s\p_z+\clA_{\bz},\Psi]=0\,, \\

3. &  \p_s\Psi=0\,.
\end{array}
\right.
\eqn
The monodromy ${\cal Y}$ of $\Psi$ is  the transformation
$$
\Psi\rar\Psi {\cal Y}, ~~{\cal Y}\in {\rm Rep}(\pi_1(\Si_{g,n})\to G).
$$
The equation 3. (\ref{2.15}) means that the monodromy is independent on the times.
The equations of motion (\ref{2.14}) for $A$ and $\bA$ are
the consistency conditions 1., 3.,  and 2.,3.
correspondingly. The consistency condition of 1. and 2. is
the flatness constraints
\beq{flmp}
(\p_{\bz}+\mu\p_z)\clA_z-\p_z\clA_{\bz}+[\clA_z,\clA_{\bz}]=\sum_a\bfS_a\de(z_a)\,.
\eq
Similar to (\ref{mcmp}), the flatness is broken at the marked points.


\subsection{Moduli space}

The form $\Om$ (\ref{esf}) is invariant under the gauge transformations
\beq{gtfb}
\clA_z\to \clA_z^f=f^{-1}\p_zf+f^{-1}\clA_zf\,,~~~
\clA_{\bz}\to \clA_{\bz}^f=f^{-1}(\p_{\bz}+\mu\p_z)f+f^{-1}\clA_{\bz}f\,,
\eq
$$
\bfS_a\to \bfS_a^f=f^{-1}(z_a,\bz_a)\bfS_af(z_a,\bz_a)\,.
$$
 They are generated by the flatness constraints (\ref{flmp}).

Up to now the equations of motion, and the linear problem are
trivial.
The meaningful equations arise after the gauge fixing
with respect to (\ref{gtfb}).
 The set  of the gauge orbits
on the constraint surface (\ref{flmp}) is the moduli space of flat connections.
By neglecting some non-generic configurations we come to
the moduli space of flat bundles
\beq{flms}
{\cal M}^{flat}_{g,n}={\rm (\ref{flmp})}/{\cal G}={\cal R}//{\cal G}.
\eq

Let us fix $\clA_{\bz}=\bL$:
\beq{2.30}
\bL=f^{-1}(\p_{\bz}+\mu\p_z) f+f^{-1}\clA_{\bz} f.
\eq
Then the dual field
\beq{2.31}
L=f^{-1}\p f+f^{-1}\clA_zf
\eq
can be found from the moment equation (\ref{flmp})
\beq{2.18}
(\p_{\bz}+\mu\p_z)L-\p_z\bL+[\bL,L]=\sum_{a=1}^n\bfS_a\de(x_a).
\eq
We preserve the notion $\bfS_a$ for the gauge transformed element of ${\cal O}_a$.
Here $L$ and $\bL$ are connections satisfying the
quasi-periodicity conditions (\ref{sect}).

The gauge fixing (\ref{2.30}) and the moment constraint (\ref{2.18})
kill almost all degrees of freedom. The fibers
$\{L,\bL,{\bfS_a}\}$ become finite-dimensional,
as well as the bundle ${\cal P}^{red}=({\cal M}^{flat}_{g,n},\clT_{g,n})$:
\beq{2.18d}
\dim\,{\cal P}^{red}=(2\dim\,G+3)(g-1)+(\dim\,\clO+1)n\,.
\eq

On ${\cal P}^{red}$ we have
\beq{red}
\Om_0=\int_{\Si_{g,n}}(\de L\wedge d\bL)+
\sum_{a=1}^n\om_{a}\,,~~~H_s=H_s(L)=\oh\int_{\Si_{g,n}}(L^2)\mu_s^0\,.
\eq
But now, due to (\ref{2.18}), the system is no long free because
$L$ depends on $\bL$, ${\bfS_a}$. Moreover, because $L$ depends explicitly
on $\mu$, the system (\ref{red}) is non-autonomous.

Let $M_s=f^{-1\p_sf}$. Then the equations of motion (\ref{2.14}) on ${\cal M}_{H}(\Si_{g,n})$
take the form
\beq{2.19}
\p_sL-\p_z M_s+[M_s,L]=0,~~s=1,\ldots,l\,,
\eq
where $M_s$ is a solution of the equation
\beq{2.20}
(\p_{\bz}+\p_z\mu)M_s-[M_s,\bL]=\p_s\bL-L\mu_s^0\,.
\eq
The equations (\ref{2.19}) are the analog of the Lax equations (\ref{leq}). The essential
difference is the presence of the operator $\p_z$ with respect to the spectral parameter.

These equations reproduce
the Schlesinger system, Elliptic Schlesinger system, multi-component
generalization of the Painlev\'{e} VI equation \cite{LO,CLOZ}.
The equations (\ref{2.19}), (\ref{2.20}) along
with (\ref{2.18}) are consistency conditions for the linear system
\beq{4.36}
\left\{
\begin{array}{ll}
  1. & [\p+L,\Psi=0]\,, \\

  2. & [\bp+\sum_st_s\mu_s^0\p+\bL,\Psi=0]\,, \\

  3. & [\p_s+M_s,\Psi]=0\,,~~(s=1,\ldots,l)\,.
\end{array}
\right.
\eq
The equations 3. provides the isomonodromy
property of the system 2., 3. with
respect to variations of the times $t_s$.
For this reason we call the nonlinear equations (\ref{2.19})
 the Hierarchy of the Isomonodromic Deformations.

 The Symplectic Hecke Correspondence can be applied to the
 monodromy preserving equations in a similar way as for the Hitchin systems.
 It allowed us to identify different descriptions of the
 Painlev\'{e} VI equation \cite{LOZ2,LZ2}.


\subsection{Scaling limit.}

Here we point out an interrelation between Hitchin integrable systems and
the monodromy preserving equations.
These interrelations were first observed at the beginning of the last century
by Garnier \cite{Ga}  and by Boutroux \cite{Bo}  and developed
 later  in numerous works \cite{FN,Kr1,Kr,Ta,Kr2,Ar}.
 Here we describe them in a very simple form in "the first approximation"
 following \cite{LO}.
 It turns out that the Hitchin systems are a some sort of quasi-classical limit of the Isomonodromy Problems. This is a sort of WKB ansatz  considered in particular
 cases in \cite{FN,Ta,No}. The next order is more complicated procedure \cite{Kr1,Kr2} leading to the so-called Whitham equations. We do not need to discuss them here.

 Introduce an analog of the Planck constant  $\ka$ and
 replace a holomorphic connection by the P.Deligne $\kappa$-connection
$\kappa\p_{z}+\clA_z$. Simultaneously, replace $\mu$ by $\mu/\ka$. Then we come to connections
 $$
 \ka\p_z+\clA_z\,,~~~\p_{\bar z}+\mu\p_z+\clA_{\bz}\,.
 $$

Consider the limit $\ka\to 0$. The symplectic form $\Om$ (\ref{esf})
is singular in this limit.
Let us  replace  the times
$$
t_s\rar t_s^0+\ka t^H_s\,,~~~(t^H_s-~{\rm Hitchin~times})
$$
and assume that the times $t_s^0, ~(s=1,\dots,l)$ are fixed.
\footnote{In the approximation procedure the times are represented by the slow variables  $t_s^0$ and the fast variables $t^H_s$. Here it is sufficient to fix
$t_s^0$.}
After this rescaling the form  (\ref{esf}) becomes regular.
The rescaling procedure means that we blow up a neighborhood
 of the fixed point $\mu(0)=\sum_st_s^0\mu_s^{(0)}$
 in ${\cal T}_{g,n}$ and consider the system in this neighborhood.
This fixed point is defined by the complex coordinates
\beq{fp}
w_0=z-\sum_st^0_s\ep_s(z,\bz)\,,~~\bar{w}_0=\bz\,,~~
 (\p_{\bar{w}_0}=\bp+\mu(0)\p)\,,~~\mu(0)=\sum_st^0_s\mu_s^{(0)}\,.
\eq
For $\ka=0$ the connection $\clA_z$ becomes the one-form $\Phi$
(the Higgs field)
$
\ka\p +\clA_z\rar \Phi\,.
$
Let
$L^0=\lim_{\ka\to 0}L$, $\bL^0=\lim_{\ka\to 0}\bL$.
Then we obtain the autonomous Hamiltonian systems with the form
$$
\Om^H=\int_{\Si_{g,n}}(\de L^0\wedge\de\bL^0)+
\sum_{a=1}^n\om_a
$$
and the commuting, time-independent quadratic integrals
$$
H_s=\oh\int_{\Si_{g,n}}(L_0^2)\mu_s^{(0)}.
$$
 The phase space turns into the cotangent bundle to the
moduli of stable holomorphic $G$-bundles over $\Si_{g,n}$.

The corresponding set of linear equations has the following form.
$$
\left\{
\begin{array}{ll}
  1. & (\ka\p+L)\Psi=0\,,
  \\
  2. & (\bp+\sum_st_s\mu_s^0\p+\bL)\Psi=0\,,
  \\
  3. & (\p_{t_s^H}+\ka\p_{t_s^0}+M_s)\Psi=0\,,~~(s=1,\ldots,l)\,.
\end{array}
\right.
$$
We consider the quasi-classical regime $\ka\to 0$
\beq{Ps}
\Psi=\phi\exp\frac{\cal S}{\ka},
\eq
where $\phi$ is a group-valued function and ${\cal S}$ is a scalar phase.
To  kill singularities we assume that
\beq{me}
(\bp+\sum_st_s\mu_s^0\p){\cal S}=0\,,~~
\frac{\p}{\p t^H_s}{\cal S}=0\,.
\eq
In the quasi-classical limit we set
\beq{SW}
 \p_{w_0}{\cal S}=\la.
\eq
Define the Baker-Akhiezer function
$$
Y=\phi\exp\sum_{s=1}^lt_s^H\frac{\p{\cal S}}{\p t^0_s}.
$$
Then instead of (\ref{4.36}) we obtain in the
limit $\ka\to 0$
$$
\left\{
\begin{array}{ll}
1.&(\la+L^0)Y=0\,,
\\
2.&(\bp_{{\bar w}_0}+\bL^0)Y=0\,,
\\
3.&(\p_s+M^0_s)Y=0,~~(s=1,\ldots,l)\,,~~(\p_s=\p_{t_s^H})\,.
\end{array}
\right.
$$
Note, that the consistency conditions for the first and the last
equations are the
standard Lax equations
\beq{2.30a}
\p_sL^0+[M_s^0,L^0]=0,
\eq
while the consistency conditions for the first and the second equations
is just the Hitchin equation, defining $L^0$
\beq{2.31a}
\p_{{\bar w}_0}L^0+[\bL^0,L^0]=2\pi i\sum_{a=1}^n\de(x_a)p_a.
\eq

It follows from (\ref{2.30a}) that the resulting Hamiltonian system is
completely integrable. The commuting integrals are
$$
H_{s,l}=\f1{d_j}\int_{\Si_{g,n}}(L_0^{d_j})\mu_{s,j}^0,~~(j=1,\ldots,l).
$$

The gauge properties of the Higgs field allows one to define the spectral
curve
$$
{\cal C}:~\det (\la +L)=0
$$
and the projection (\ref{pro}).


\section{Gauge theory description}
\setcounter{equation}{0}

Here we present results of Hitchin \cite{Hi1}. In this paper he described
two-dimensional reductions of the self-duality equations in dimension four.
We shall focus on the fact that there are two descriptions of the moduli space  of their solutions. They are the phase spaces of integrable systems or the phase spaces of  monodromy preserving equations, defined above.

\subsection{2-d self-dual equations}

Consider the self-duality equation in the Yang-Mills theory
on $\mR^4$ with coordinates
$\bfx=(x_0,x_1,x_2,x_3)$
with a gauge group $G_c$, where $G_c$ is a simple compact Lie group
\beq{sd4}
F=\star F\,.
\eq
Here $\star$ is the Hodge operator and the curvature is a two-form
$F(\bfA)\in \Om^{(2)}(\mR^4,\gg_c) $
$F_{ij}=[\nabla_i,\nabla_j]$ or $F(\bfA)=d\bfA+\bfA^2$ taking values
 in the Lie algebra $\gg_c$.
If $\bfx=(x_0,x_1,x_2,x_3)$ are
 coordinates on $\mR^4$ then (\ref{sd4}) takes the form
 \beq{sd}
\left\{
\begin{array}{c}
F_{01}=F_{23}\\
F_{02}=F_{31}\\
F_{03}=F_{12}
\end{array}
\right.
\eq

Assume that $A_j$ depend only on $(x_1,x_2)$. This means that
the fields are invariant under the shifts in directions $x_0,x_3$.
Then $(A_0,A_3)$ become adjoint-valued
one-forms which we denote  as $(\phi_1,\phi_2)$.
They are called  the Higgs fields.
\footnote{
The Higgs fields in particle physics are scalar fields, but not
one-forms. In fact, $(\phi_1,\phi_2)$  are the Higgs field
in the SUSY $\clN=4$ Yang-Mills theory. They become one-forms after
some topological twist \cite{KW}.}
 The self-dual equations on the plane $\mR^2=(x_1,x_2)$ take the form
\beq{sd1}
F_{12}=[\phi_1,\phi_2]\,,
\eq
\beq{sd2}
[\nabla_1,\phi_1]= [\phi_2,\nabla_2]\,,
\eq
\beq{sd3}
[\nabla_1,\phi_2]=[\nabla_2,\phi_1]\,.
\eq

Introduce  complex coordinates
$z=x_1+ix_2\,,$ $\,\bz=x_1-ix_2$ and let
$\,d'=\nabla_z$, $\,d''=\nabla_{\bz}$. Consider
the complexification
$$
\left\{
\begin{array}{c}
\Phi_z=\oh(\phi_1-i\phi_2)dz\in\Om^{(1,0)}(\mR^2,\gg_c)\,,\\
 \Phi_{\bz}=\oh(\phi_1+i\phi_2)d\bz\in\Om^{(0,1)}(\mR^2,\gg_c)\,.
 \end{array}
 \right.
$$
$$
\left\{
\begin{array}{c}
A_z=\oh(A_1-iA_2)\\
 A_{\bz}=\oh(A_1+iA_2)\,,
 \end{array}
 \right.
$$
In terms of  fields
\beq{W}
\clW=(A\,,A_{\bz}\,,\Phi_z\,,\Phi_{\bz})
\eq
 (\ref{sd1}) -- (\ref{sd3}) can be rewritten in the coordinate invariant way:
\beq{he1}
\left\{
\begin{array}{ll}
1.\,& F+[\Phi_z,\Phi_{\bz}]=0\,,\\
2.\,& d_{A_{\bz}}\Phi_z=0\,,\\
3.\,& d_{A_{z}}\Phi_{\bz}=0\,,
\end{array}
\right.
\eq
where $[\Phi_z,\Phi_{\bz}]=\Phi_z\Phi_{\bz}+\Phi_{\bz}\Phi_z\,$.

Equations (\ref{he1}) are conformal invariant and thereby can be defined
on a complex curve $\Si_g$ with local coordinates $(z,\bz)$.
Let $\Om^{(j,k)}(\Si_g,\gg_c)$ be $(j,k)$-forms on  $\Si_g$ taking values in
ad$(\gg_c)$. Then
$$
\Phi_z\in\Om^{(1,0)}(\Si_g,\gg_c)\,,~~\Phi_{\bz}\in\Om^{(0,1)}(\Si_g,\gg_c)\,,
$$
$$
d_{A_{\bz}}\,:\,\Om^{(j,k)}(\Si_g,\gg_c)\to\Om^{(j,k+1)}(\Si_g,\gg_c)\,.
$$
The self-duality equations (\ref{he1}) on  $\Si_g$
are called \emph{the Hitchin equations}.

In fact, instead of (\ref{he1}) we will consider further a modified system
\beq{he}
\left\{
\begin{array}{ll}
1.\,& F-[\Phi_z,\Phi_{\bz}]=0\,,\\
2.\,&d_{A_{\bz}} \Phi_z=0\,,\\
3.\,& d_{A_{z}}\Phi_{\bz}=0\,.
\end{array}
\right.
\eq
It comes from the self-duality on $\mR^4$ with a metric of signature $(2,2)$.


\subsection{The moduli space}

The system (\ref{he}) is invariant with respect to the gauge transformations
\beq{gt3}
\clG_c=\{f\in\Om^{0}(\Si_g,G_c)\}\,,
\eq
\beq{gt1}
\Phi_z\to f^{-1}\Phi_z f\,,~~\Phi_{\bz}\to f^{-1}\Phi_{\bz} f\,,
\eq
\beq{gt2}
d_{A_{\bz}}\to f^{-1}d_{A_{\bz}}f\,,~~d_{A_{z}}\to f^{-1}d_{A_{z}}f\,.
\eq
If $(A\,,A_{\bz}\,,\Phi_z\,,\Phi_{\bz})$ are solutions of  (\ref{he}), then
the transformed fields are also solutions.
Define the moduli space of solutions of (\ref{he}) as a quotient under the gauge
group action
\beq{ms}
\clM_H(\Si_g)={\rm solutions~of~(\ref{he})}/\clG_c\,.
\eq
Dimension of this space is
\beq{dhms}
\dim\,(\clM_H(\Si_g))=2(g-1)\dim\,(G_c)\,.
\eq

For generic configurations  of the fields the space $\clM_H(\Si_g)$ has two equivalent description:

\textbf{I)} Consider the pair $(\Phi=\Phi_z\,,\bA= A_{\bz})$ taking values in the complex algebra $\gg$.
Assume that they satisfy the second equation in (\ref{gt3}). The complex gauge
group $\clG=\{f\in\Om^{0}(\Si_g,G)\}$ transforms its solutions into another solutions.
Taking the quotient of solutions with respect to this action we come to
 the moduli space of the Higgs bundles, described in Section 4.
It was found in \cite{Hi2} (see, also, \cite{KW}) that for generic pairs $(\Phi\,,\bA)$
\beq{hbms}
\clM^{G_c}_H(\Si)\sim T^*\clM_g
\eq

\textbf{II)} Define the complex valued connection
$\clA_z=A_z+i\Phi_z\,,$ $\,\clA_{\bz}=A_{\bz}-i\Phi_{\bz}$.
The first equation in (\ref{he}) becomes the flatness condition for the curvature $\clF=d\clA+\clA\wedge\clA$
\beq{flc}
\clF=0\,.
\eq
Evidently, it is invariant under the action of the complex gauge group $\clG$.
Then generic flat bundles describe the Hitchin moduli space as the quotient
\beq{flms1}
\clM^{G_c}_H(\Si_g)\sim \clM_g^{flat}=(\clF=0)/\clG\,.
\eq
This moduli space has been defined (\ref{flms}).

These two description is based on the hyperkahler structure of $\clW$ (\ref{W})
\cite{Hi2,KW}. It implies an existence on $\clW$ three complex structures $I,J$, $K=IJ$
and three symplectic forms $\Om_I$, $\Om_J$ and $\Om_K$. They are $(2,0)$ forms in the corresponding complex structures.
The first constructions is based on symplectic reduction with respect to the form $\Om_I$, which coincides with (\ref{sfh}). The second constructions is the result of symplectic reduction with respect to the form $\Om_J$ (\ref{cla}).

As a result due to this procedure we identify the phase spaces of the Hitchin
integrable systems and the monodromy preserving equations. In fact, these two
moduli spaces are isomorphic as real manifolds but not as complex manifolds.


\section{Bogomolny equation}

\setcounter{equation}{0}

Here following \cite{KW} we interpret the modification in terms of monopole configurations.
The relevant  monopole configurations are singular solutions of the Bogomolny equation. These configurations of fields correspond to the t'Hooft operators
in the underlying Yang-Mills theory. The present treatment is based on our paper
\cite{LOZ3}, where we considered the modifications for flat bundles.

\subsection{Definition}
Let $W=\mR\times\Si_g$. Consider a $G$ bundle $E$ over $W$
equipped with the curvature $F$. We assume that
  $G$ is a complex group and
$E$ is an adjoint bundle. It implies that $F$ takes values in the
complex Lie algebra $\gg$.
Let $\phi$
be a zero form on $W$ $\phi\in\Om^0(W,\gg)$.

 The Bogomolny equation on $W$ is a three-dimensional reduction of the self-duality
 equation (\ref{sd}).
It takes the form
\beq{q1}
F=*D\phi\,.
\eq
Here $*$ is the Hodge operator on $W$  with respect to the metric $ds^2$ on $W$.
In local coordinates $(z, \bz)$ on $\Sigma_{g,n}$ and $y$ on the real line
 $ds^2=h|dz|^2+dy^2$,
where $h(z,\bz)|dz|^2$ is a metric on $\Si_g$.
Then the Hodge operator is defined as
$$
\star dy=\oh i hdz\wedge d\bz\,,~
\star dz= -i dz\wedge dy\,,~
\star d\bz= i d\bz\wedge dy\,,
$$
In local coordinates (\ref{q1}) takes the form
\beq{q2}
\left\{
\begin{array}{ll}
1.&\p_{z} A_{\bz}-\p_{\bz} A_z+[A_z,A_{\bz}]=\frac{ih(z,\bz)}{2}\left(\p_y\phi+[A_y,\phi]\right)\,,
\\
2.&\p_y A_z-\p_{z} A_y+[A_y,A_z]=i(\p_{z}\phi+[A_z,\phi])\,,
\\
3.&\p_yA_{\bz}-\p_{\bz} A_y+[A_y,A_{\bz}]=-i(\p_{\bz}\phi+[A_{\bz},\phi])\,.
\end{array}
\right.
\eq

 A singular monopole solution of this equation is obtained  in the following way.\\
 Let $\ti{W}=(W\setminus \vec{x}^0=(y=0,z=z_0))$.
The Bianchi identity $ DF=0$ on $\ti{W}$ implies that
$\phi$ can be identified with the Green function for the operator $\star D\star D$
\beq{GF}
\star D\star D\phi=\ga\de(\vec{x}-\vec{x}^0)\,,~~\ga\in\gH\,.
\eq
We take $\ga=\sum_{j=1}^lm_j\varpi_j^\vee$ from the coweight lattice $P^\vee$ (\ref{cwl}) as in the modification procedure (\ref{mtm}).
We call the Green function a monopole with charges $(m_1,m_2,\ldots,m_n)$.
 We explain below this choice of the coefficient in front of the delta-function.
This equation means that $\phi$ is singular at
$\vec{x}^0$.


\subsection{Boundary conditions and gauge symmetry}.

In what follows
 we assume that
 \beq{bcb}
 \lim_{y\to \pm\infty}(\p_y\phi)=0\,.
 \eq
It is the Neumann boundary conditions for the Higgs field, while the gauge
fields are unspecified.
 Let $E_\pm$ be restrictions of $E$ to the bundles over $\Si_g$
 on the "left end" and "right end" of
$W\,:\, y\to \pm\infty$. These bundles are flat.
 This fact follows from 1.(\ref{q2}), where the gauge fixing $A_y=0$ is assumed.
The Bogomolny equation defines a transformation $E_-\to E_+$.
It was proved in \cite{KW} that
  in  absence of  the source $\ga=0$ in (\ref{GF}) the only solutions of (\ref{q1})
with these boundary conditions  are $F\equiv 0\,,$ $\phi\equiv 0$.
In general, under the action of the source
the characteristic classes of bundles are changed under these
transformations. We will find that it is the modification of the type
$(m_1,m_2,\ldots,m_n)$.

The system (\ref{q2}) is invariant with respect to the gauge group $\clG$ action:
\beq{q3}
\begin{array}{c}
A_z\rightarrow hA_zh^{-1}+\p_{z} hh^{-1}\
, \ \ \
A_{\bz}\rightarrow hA_{\bz} h^{-1}+\p_{\bz} hh^{-1}\,, \ \ \
A_y\rightarrow hA_yh^{-1}+\p_y hh^{-1}
\\
\phi\rightarrow h\phi h^{-1}\,,
\end{array}
\eq
where $h\in\clG$ is a smooth map $W\to G$. To preserve the r.h.s in (\ref{GF})
it should satisfy the condition $[h(\vec{x}^0),\ga]=0$.

Since the gauge fields for $y=\pm\infty$ are unspecified and only flat we can act on them by boundary values of the gauge group $\clG|_{y=\pm\infty}=\clG_\pm$.
Then $\clM^\pm=\{E_\pm\}/\clG_\pm=\clM_H^\pm(\Si_{g,n})$ are the moduli spaces of flat bundles (see (\ref{flms}), (\ref{flms1})).
 In this way a monopole solution put in a correspondence
 two moduli spaces $\clM_H^\pm(\Si_{g,n})$. But Bogomolny
equation tells us more. It describes an evolution from one type of system to another.

It is possible to generalize (\ref{GF}) and consider multi-monopole sources $\sum_a\ga_a\de(\vec{x}-\vec{x}_a^0)$  in the
r.h.s. .
This generalization will correspond to modifications in a few points of $\Si_{g,n}$.


\subsection{Gauge fixing}
Consider (\ref{q2}) on $\widetilde{W}$.
Choose a gauge fixing conditions as: $A_{\bz}=0$. Holomorphic in $z$ functions $h=h(y,z)$ preserve this gauge. Then
\beq{q4}
\left\{
\begin{array}{l}
-\p_{\bz} A_z=\frac{ig}{2}\left(\p_y\phi+[A_y,\phi]\right)\,,
\\
\p_y A_z-\p_{z} A_y+[A_z,A_y]=i(\p_{z}\phi+[A_z,\phi])\,,
\\
\p_{\bz} A_y=i\p_{\bz}\phi\,.
\end{array}
\right.
\eq
The last equation means that $A_y-i\phi$ is holomorphic. It follows from (\ref{q3})
that the gauge transformation of this function is
$$
A_y-i\phi\rightarrow h(A_y-i\phi)h^{-1}+\p_y hh^{-1}
$$
Thus, we can keep $A_y=i\phi$ by using holomorphic and $y$-independent part of the gauge group $(\p_yh=0)$.
Finally, we come to the system
\beq{q5}
\left\{
\begin{array}{ll}
1.\,&\p_{\bz} A_z=-\frac{ih}{2}\p_y\phi\,,
\\
2.\,&\p_y A_z-2i\p_{z} \phi+2i[A_z,\phi]=0\,,
\\
3.\,& A_y=i\phi\,,
\\
4.\,& A_{\bz}=0\,.
\end{array}
\right.
\eq
Two upper equations from (\ref{q5}) lead to the Laplace type equation
$$
\p^2_y\phi+\frac{4}{h}(\p_{z}\p_{\bz}\phi+\p_{\bz}[A_z,\phi])=0\,,
$$
or on $W$
\beq{q61}
\p^2_y\phi+\frac{4}{h}(\p_{z}\p_{\bz}\phi+\p_{\bz}[A_z,\phi])=M\de(y,z_0)\,.
\eq


\subsection{Scalar case.}

In scalar case (\ref{q61}) is simplified
\beq{q7a}
\p^2_y\phi+\frac{4}{h}\p_{z}\p_{\bz}\phi=c\de(y=0,z_0)\,.
\eq
where for the time being the value $c$ is not specified.

Consider (\ref{q7a}) on $\mR\times\mC$ with the global
 coordinates $(z,\bz,y)$ and $h=1$.
Then
\beq{q10}
\phi=c\frac{1}{\sqrt{y^2+z\bz}}
\eq
satisfies (\ref{bcb}). It follows from
1.(\ref{q5}) that
\beq{q101}
\left\{
\begin{array}{l}
A_z(z,\bar{z},y)=A_z^+(z,\bar{z},y),\ y>0\ and\ y=0, z\neq 0\,,
\\
A_z(z,\bar{z},y)=A_z^-(z,\bar{z},y),\ y<0\,,
\end{array}
\right.
\eq
where
$$
\begin{array}{c}
A_z^+(z,\bar{z},y)=-ic\left(\frac{1}{z}\frac{y}{\sqrt{y^2+z\bz}}-\frac{1}{z}\right)
+const\,,
\\
A_z^-(z,\bar{z},y)=-ic\left(\frac{1}{z}\frac{y}{\sqrt{y^2+z\bz}}+\frac{1}{z}\right)
+const\,,
\end{array}
$$
and $A_z(z,\bar{z},y)$ is a connection on the line bundle $\clL$ over $\mR\times\mC$.
The connection has a jump $-2ic\frac{1}{z}$ at $y=0$. To deal with smooth connections
we compensate the jump by a holomorphic  gauge transform that locally
near $\vec x_0$ has the form $h\sim z^m$. Here $m$ should be integer,
because $h$ is a holomorphic function. Notice that all holomorphic line bundles over  $S^2$ are known to be
 $\mathcal{O}(m)$-bundles, $m\in\mZ$.
Thus, we have $c=i\frac{m}{2}\,$,  $m\in\mathbb{Z}$. This usually referred as
a quantization of the monopole charge. In fact the constant $c$ contains factor $4\pi$ (area of a unit sphere)
which yields a proper normalization of delta-function and appears in the Gauss's law.


\subsection{Modification}.

Assume that
near the singular point $\vec x^0$ the Higgs field $\phi$ can be taken from
the Cartan subalgebra.
Using the solution (\ref{q10}) for a line bundle we write in in the form
\beq{phi}
\phi|_{\vec{x}\to\vec x^0}\sim\frac{i\ga}{2\sqrt{y^2+z\bz}}\in\gH\,.
\eq
It follows from (\ref{q101}) that $A_z$ undergoes a
discontinuous jump at $y=0$
\beq{awga}
A^+_{z}-A^-_{z}=\frac{i\ga}z\,,~~~\ga\in\gH\,.
\eq
To get rid of the singularity of $A$ at $z=0$,
as in the abelian case, one can perform the
singular gauge transform $\Xi$ that behaves near $\vec x^0$ as
\beq{modi}
\Xi\sim z^\ga\in\clH\,.,~~~(\Xi=\exp\,(2\pi i\ga\ln z))\,.
\eq
Assume that $\ga$ belongs to the coweight lattice $\ga\in P^\vee$ as in  (\ref{mtm}).
It means that $\Xi$ is the modification of type $\ga=\sum_{j=1}^lm_j\varpi_j^\vee$.
As it was explained before, the modified bundle $\ti E$ can not be lifted to a $\bar G$ bundle. Thus, if the monopole charge belong to $P^\vee$ then the solution of the Bogomolny equation defines the modification of the initial flat bundle $E_-$ to the modified
bundle $E_+=\ti E$.
On the other hand, if $\ga$ belongs to
the coroot lattice $Q^\vee$ (\ref{cjrl}), then  there is no obstruction to
lift $\ti E$ to an $\bar G$ bundle. In other words, the monopoles with charges from
the coroot lattices do not change the topological type of bundles. Therefore, monopoles with nontrivial charges are classified by the quotient $P^\vee/Q^\vee\sim\clZ(\bG)$ (\ref{center}). In this way there are no nontrivial monopoles for the
groups $G_2$, $F_4$ and $E_8$.
From field-theoretical point of view the modification (\ref{modi})
 it is an action of the t'Hooft operator.

In summary, the monopole solutions  of the Bogomolny equation
allows one to relate them to the symplectic Hecke
correspondence in the monodromy preserving equations. Using the correspondence
between the monodromy preserving equations and the Hitchin integrable systems we
relate the same  configurations  to the symplectic Hecke
correspondence in the Hitchin integrable  systems.


\section{Twisted $\clN=4$ Super Yang-Mills Theory and Integrable systems.}

In \cite{KW} Kapustin and Witten described the Langlands program in terms of the
 $\clN=4$, $d=4$ SUSY  Yang-Mills Theory  with compact gauge group $G_c$.
By  certain twists the theory becomes topological and $\mR^4$ can be replaced by
a theory  on $C\times \Si$, where
$\Si$ will play the role of the basic spectral curve introduced above, and $C$
is a Riemann surface with boundaries.\footnote{Because $\Si$ is a standard notation for basic spectral curves, our notations are opposite to \cite{KW}, where $\Si$ is denoted as $C$ and $C$ as $\Si$.}
 In particular, $C$ can be taken as $\mR^2$. It
includes time $t$ and $y$ variables that was used in the Bogomolny equation.
One of the crucial points is that in the limit when $\Si$ becomes small compared
with $C$ the effective theory is described by a topological sigma-model $C\to\clM_H(\Si)$.

In fact, there is a family of twists in  $\clN=4$ theories. For a one value of the
twist parameter the theory can be reduced to the Bogomolny equation in three
dimension and to the Hitchin equations (\ref{he1}). The mention above sigma-model
is the A-model with the target space
$\clM^{G_c}_H(\Si_{g})$. It is described in the form (\ref{hbms}), i.e. the
phase space of  the Hitchin integrable system, related to the group $G$

For another specific value of the twist parameter one should consider
the dual gauge group $^LG_c$ and the two-dimensional sigma-model is the B-model
with the target space $\clM^{^LG_c}_H(\Si_{g})$ (\ref{flms}).
The supersymmetry condition is equivalent to flatness of the complexified connections
 (\ref{flc}). It is the phase space for the monodromy preserving equations related
 to the group $^LG$.
 The S-duality in the Yang-Mills Theory in $d=4$ becomes the mirror symmetry between
 A-models with the target space $\clM^{G_c}_H(\Si_{g,n})$   and B-models
  with the target space $\clM^{^LG_c}_H(\Si_{g})$.
 In this way the Hitchin systems are dual to the  monodromy preserving equations
  related to the dual groups (see Table 2 in Appendix).

 In the A-model the natural  operators  are the t'Hooft operators.
 They correspond to singular configurations of the gauge field
  on a line or on a loop in $d=4$. In the abelian case
 $G_c=U(1)$
 they are defined as Dirac monopole configuration  (\ref{q101}).
 In the non-abelian case with a compact gauge $G_c$ we define 
a singular gauge transform  using the map $t_c\,:U(1)\to G_c$.
   Since  $t_c$ is defined  up to conjugations consider
 the map $t_c\,:U(1)\to T_c$ in a Cartan torus $T_c\subset G_c$.
 Continue  $t_c$ to the holomorphic map of $\mC^*$ to the Cartan subgroup $\mC^*\to \clH_G$ of the complex group $G$.
  $t_c$ a co-character of  $G$ (\ref{coch}). In this way we come to  the modification
 (\ref{modi}). This construction establish a connection between the  t'Hooft operators in the Yang-Mills theories and the Symplectic Hecke Correspondence in the Hitchin Systems.

The S-dual to the t'Hooft operators are the Wilson operators in Yang-Mills theory
with the group $^LG_c$ \cite{MO,Ka}. We will not discuss here their implications in
the theory of Integrable systems. 

Another important class of singular operators
that related to this theory - singular operators on two-dimensional surfaces
transversal to $\Si_{g}$
\cite {GW,GW1}. Some special types of these operators correspond to coadjoint
orbits at the marked points of $\Si_{g}$.
They are related to the spin variables at the marked points. It was explained in \cite{GW} that in a neighborhood
of a marked point the Hitchin equations can be written in the form of the Nahm
equations. In this way the part of the phase spaces related to the spin variables
can be described in terms of the moduli space of solutions of the Nahm equation.

Summarizing we repeat the correspondence between objects in gauge theories
and classical integrable systems.

$$
\begin{array}{ll}
 {\rm {\bf Yang-Mills ~theory}} &  {\rm{\bf Integrable~systems }}\\
      &   \\
   Gauge~group  &  Hidden ~symmetry  \\
  2d~ surface &  Basic~ spectral~ curve \\
     Higgs ~fields  & Lax~ operators\\
 t'Hooft ~operators  &  Modification  \\
    Rigid~surface~operators  & Spin ~variables  \\
    Moduli~parameters & Coordinates~ of~particles\\
      Supersymmetry ~conditions  & Hitchin~equations \\
      S-duality & Duality ~between~ Hitchin~systems\\
            & ~and~monodromy~preserving~equations\\
            & for~dual~groups
\end{array}
$$

\section{Appendix.
 A piece of group theory \cite{OV,Bo1}}
\setcounter{equation}{0}
\def\theequation{A.\arabic{equation}}

 Let $\gg$ be a Lie algebra of a simple complex Lie group $G$,
$\gH$ its Cartan subalgebra, $(\dim\,( \gH)$ is a rank of $G$).
Let  $\gH^*$ be a dual to $\gH$ space, and
  $\lan~,~\ran$ is a pairing between $\gH$ and  $\gH^*$.
A finite system of vectors $R=\{\al\}$ in $\gH^*$ is called a root system, if\\
1. $R$ generates $\gH^*$;\\
2. For any $\al\in R$ there exists a coroot $\al^\vee\in\gH$ such that
$\lan\al,\al^\vee\ran=2$ and the reflection in $\gH^*$
\beq{ref}
s_\al\,:~x\mapsto x-\lan x,\al^\vee\ran\al
\eq
preserving $R$;\\
3.$\lan\be,\al^\vee\ran\in\mZ$ for any $\be\in R$;\\
4. For $\al\in R$ $\,n\al\in R$ iff $n=\pm 1$.

The dual  system $R^\vee=\{\al^\vee\}$ is the root system in $\gH$.
The group of automorphisms of $\gH^*$ generated by reflections (\ref{ref}) is the Weyl group $W(R)$. The groups $W(R)$ and $W(R^\vee)$ are isomorphic.

Define a basis $\Pi=(\al_1,\ldots,\al_l)$ of simple roots in $R$ such that any $\al\in R$ is decomposed in this basis as
\beq{rde}
\al=\sum_{j=1}^ln^\al_j\al_j\,, ~~m_j\in\mZ\,,
\eq
and all $m_j$ are positive (in this case $\al$ is a positive root), or negative
($\al$ is a negative root). In other words the root system is an union of positive and negative roots $R=R^+\cup R^-$.

Let $S^W$ be an algebra of polynomials on $\gH$ invariant with respect to $W$-action.
There exists a basis in $S^W$ of $l$ homogeneous polynomials of degrees
$d_1=2, d_2,\ldots,d_l$. The degrees are unequally defined by the root system $R$.
The number of roots can be read off from the degrees
\beq{nro}
\sharp\,R=2\sum_{i=1}^l(d_i-1)\,.
\eq

Let
$Q=\sum_{j=1}^nn_j\al_j\,,~$ $(n_j\in\mZ\,,\,\al_j\in\Pi)$ be  a root lattice in  $\gH^*$. The simple coroots $\Pi^\vee=(\al^\vee_1,\ldots,\al^\vee_l)$ generate the
coroot lattice in $\gH$
\beq{cjrl}
Q^\vee=\sum_{j=1}^nn_j\al^\vee_j\subset\gH\,.
\eq

The fundamental weights $\varpi_j\in\gH^*\,,$ $\,(j=1,\ldots,n)$
 form a dual basis in $\gH^*$
$\lan\varpi_j,\al^\vee_k\ran=\de_{jk}$. They generate the weight lattice
$P=\sum_{j=1}^lm_j\varpi_j\subset\gH^*$.
In other words, $P$ is dual to the coroot lattice $Q^\vee$.
Similarly, the coweight lattice
\beq{cwl}
 P^\vee=\sum_{j=1}^lm_j\varpi^\vee_j\,,~~m_j\in\mZ\,,~~~
\lan\varpi^\vee_j,\al_k\ran=\de_{jk}\,.
\eq
This lattice is dual to the root lattice $Q$.

\bigskip
\noindent
\emph{\textbf{Root decomposition}}\\
The algebra $\gg$ has the root decomposition
\beq{CD}
\gg=\gH+\gL\,,~~\gL=\sum_{\be\in R}c_\be E_\be\,,  ~~c_\be\in\mC\,.
\eq
Here $c_\be E_\be$ are root subspaces. It follows from (\ref{nro}) that
\beq{dig}
\dim\,\gg=\sum_{i=1}^l(2d_i-1)\,.
\eq

The Chevalley basis in $\gg$ is generated by
\beq{CBA}
\{E_{\be_j}\,,~\be_j\in R\,,~~H_{\al_k}=\al_k^\vee\in\Pi^\vee\}\,.
\eq
To construct the basis consider $H_{\al_k}=\al_k^\vee$, $\,E_{\pm\al_j}$,
$(\al_j\in\Pi)$. They generate the basis for the whole algebra.
The commutation relations for these generators take the form
\beq{crsr}
\begin{array}{ll}
  (i) & [H_{\al_k},H_{\al_j}]=0\,, \\
  (ii) & [E_{\al_k},-E_{\al_k}]=H_{\al_k}\,, \\
  (iii) & [H_{\al_k},E_{\pm\al_j}]=\pm a_{jk}E_{\pm\al_j}\,, \\
  (iv) & (ad(E_{\pm\al_i})^{1-a_{ji}}(E_{\pm\al_j})=0\,,~~(i\neq j)\,.
\end{array}
\eq

Let $B$ be a Borel subgroup of $G$. It is generated
by Cartan subgroup of $G$ and by negative root subspaces $\exp\,(\sum_{\al\in R^-}E_\al)$. The coset space $Fl=G/B$ is called\emph{ the flag variety}.
It has dimension (see (\ref{nro}))
\beq{fld}
\dim\,Fl=\sum_{j=1}^l(d_j-1)\,.
\eq
  The coadjoint orbits
\beq{co}
\clO=\{Ad^*_gS_0\,|\,g\in G\,,~S_0~{\rm is~a fixed~element~of~}\gg^*\}\,.
\eq
is a generalization of a cotangent bundle to the flag varieties,
\footnote{It is a cotangent bundle if $S_0$ is a Jordan element. If $S_0$ is
semisimple,then $\clO$ is the so-called torsor over $Fl$.} and for generic orbits
\beq{dio}
\dim\,\clO=2\sum_{j=1}^l(d_j-1)\,.
\eq

\bigskip

\noindent
\emph{\textbf{Characters and cocharacters.}}\\
Let $\clH$ be a Cartan subgroup $\clH\subset G$.
 Define the group of characters
\footnote{
The holomorphic maps of $\clH$ to $\mC^*$ such that $\chi(h_1h_2)=\chi(h_1)\chi(h_2)$ for $h_1,h_2\in\clH$.}
\beq{cha}
\G( G)=\{\chi\,:\,\clH\to\mC^*\}\,.
\eq
This group can be identified with a lattice group in $\gH^*$ as follows.
Let $\bfx=(x_1,x_2,\ldots,x_l)$ be an element of $\gH$, and
$\exp\,2\pi i\bfx\in\clH$. Define $\ga\in \gH^*$ such that
\beq{char1}
\chi=\exp 2\pi i \lan\ga,\bfx\ran\in\G( G)\,.
\eq
This map is well defined only for discrete values of $\ga$. For example, for
$\bG={\rm SL}(2,\mC)$ $\clH=\di(\exp\,2\pi ix,\exp\,-2\pi ix)$, where $0\leq x<1$. Then
$\chi=\exp\,2\pi ikx$, where $k\in\mZ$. For $G^{ad}={\rm SL}(2,\mC)/\di(-1,-1)$
$\chi=\exp\,2\pi ikx$, where $k\in 2\mZ$.

Let $\bar G$ be the universal covering group of $G$ and $G^{ad}$ is the
adjoint group ($G^{ad}=\bar G/\clZ(\bar G)\,$,
$\clZ(\bar G)$ is a center of $\bar G$). Let $\mu_l$ be a subgroup of $\clZ(\bar G)$
such that $G$, or in more details, $G_l$ is a factor-group
\beq{glf}
G=G_l=\bG/\mu_l\,.
\eq
 It means that
nontrivial groups $G\nsim G^{ad},\bG$ can arise only in $A_{n-1}$ ($n$ is not prime)
and $D_n$ cases.

Characters of $\bar G$ and $G^{ad}$ are
\beq{char}
\G(\bar G)=P\,,~~\G(G_{ad})=Q\,,
\eq
and $\G(G^{ad})\subseteq\G(G)\subseteq\G(\bar G)$.
The fundamental weights $\varpi_k\,$ $(k=1,\ldots,n)\,$
(simple roots $\al_k$) form a basis in $\G(\bar G)$
($\G(G_{ad})$). Let $p$ be a divisor of $ord\,(\clZ(\bar G))$
such that $l=ord\,(\clZ(\bar G))/p$.
Then the lattice $\G(G)$ is defined as
\beq{cg}
\G(G)=Q+\varpi\,,~~ p\varpi\in Q\,.
\eq

Define the dual groups   of co-characters  $t(G)=\G^*(G)$
as holomorphic maps
\beq{coch}
t(G)=\{\mC^*\to\clH\}\,.
\eq
In another way $t(G)$ is defined as the kernel of the exponential map
$\exp\,:\gH \to \clH$
\beq{tb1}
t(G)=\{\bfx\in \gH\,|\,\exp\,(2\pi i\bfx)=1\}\,.
\eq
We find from (\ref{char}) that  the groups $t(\bar G)$ and $t(G^{ad})$
 are the coroot and the coweight lattices
\beq{tb}
t(\bar G)=Q^\vee\,,~~t(G_{ad})=P^\vee\,.
\eq
Thus a generic element of $t(G)$ takes the form
\beq{cocharc}
z^\ga=\exp\,2\pi i \ga\ln z\in\clH\,,~~\ga\in P^\vee~{\rm or}~Q^\vee\,.
\eq
In the intermediate case we have $t(\bar G)\subseteq t(G)\subseteq t(G^{ad})$.
It follows from (\ref{cg}) that
\beq{coch1}
t(G_l)=Q^\vee+\varpi^\vee\,,~~l\varpi^\vee\in Q^\vee\,.
\eq

It follows from (\ref{crsr}), that $Ad_{\zeta=\exp\,2\pi i\ga}(\bfx)=\bfx$ for $\ga\in P^\vee$. On the other hand $\zeta$ is a nontrivial element in $\bG$ and $\zeta=1$ in $G^{ad}$ (see (\ref{tb1}) and (\ref{tb})). Therefore,
\beq{center1}
\zeta=\exp\,2\pi i\ga\in\clZ(\bar G)~{\rm for~}\ga\in P^\vee\,.
\eq
In fact,
\beq{center}
\clZ(\bar G)=P^\vee/t(\bar G)\sim P^\vee/Q^\vee\,.
\eq
In general, the center $\clZ(G)$ of $G$ is
\beq{cG}
\clZ(G)\sim P^\vee/t(G)=\mu_p\,.
\eq

All groups except
of $G_2$, $F_4$ and $E_8$  have nontrivial centers.
\begin{center}
\texttt{Table 1\\
Centers of universal covering groups
}\\
($\mu_N=\mZ/N\mZ$)

\vspace{3mm}

\begin{tabular}{|c|c|c| }
  \hline
   $\bG$ &Lie $(\bar{G})$ & $\clZ(\bar{G})$ \\
 \hline
SL$(n,\mC)$ &  $A_{n-1}$ & $\mu_n$  \\
Spin$_{2n+1}(\mC)$&  $B_n$ & $\mu_2$  \\
Sp$_n(\mC)$&  $C_n$ & $\mu_2$   \\
Spin$_{4n}(\mC) $&  $D_{2n}$& $\mu_2\oplus\mu_2$     \\
Spin$_{4n+2}(\mC) $&  $D_{2n+1}$ & $\mu_4$   \\
$E_6(\mC)$ &  $E_6$ & $\mu_3$   \\
$E_7(\mC)$ &  $E_7$ & $\mu_2$   \\
  \hline
\end{tabular}
\end{center}
\vspace{5mm}

The root system $R$ of a Lie algebra $\gg=Lie(G)$, the group of characters $\G(G)$ and cocharacters $t(G)$ are called \emph{the root data}.
\emph{A Langlands dual} to $G$ group $^LG$ is defined by  the root data $R^\vee$
and
\beq{ldg}
t(^LG)\sim\G(G)\,,~~~\G(^LG)\sim t(G)\,.
\eq
If $G_l=\bG/\mu_l$, then $^LG_l=G_p=\bG/\mu_p$, where $pl=ord\,(\clZ(\bar G))$.
In  particular, $^L\bG=G^{ad}$.

\begin{center}
\texttt{Table 2\\
Duality in simple groups
}\\
\vspace{3mm}

\begin{tabular}{|c|c|c|}
  \hline
  Root system & $G$ & $^LG$ \\
  \hline
  $a_n$, $N=n+1=pl$ & $G_l=\SLN/\mu_l$ & $G_p=\SLN/\mu_p$ \\
  $b_n$ & Spin$(2n+1)$ & Sp$(n)/\mu_2$ \\
  $c_n$ &  Sp$(n)$ & SO$(2n+1)$ \\
  $d_{2l+1}$ & Spin$(4l+2)$ & SO$(4l+2)/\mu_2$ \\
   &  SO$(4l+2)$ & SO$(4l+2)$ \\
  $d_{2l}$ & Spin$(4l)$ & SO$(4l)/\mu_2$  \\
   & SO$(4l)$ &  SO$(4l)$ \\
  $d_{4l}$ & Spin$^L(8l)$ & Spin$^L(8l)$ \\
   & Spin$^R(8l)$ & Spin$^R(8l)$ \\
  $d_{4l+2}$  & Spin$^L(8m+4)$ & Spin$^R(8m+4)$ \\
  $g_2$    &  $G_2$ &  $G_2$\\
  $f_4$    &  $F_4$ &  $F_4$\\
   $e_6$ & $E_6$ & $E_6/\mu_3$ \\
  $e_7$ & $E_7$ & $E_7/\mu_2$ \\
    $e_8$ & $E_8$ & $E_8$ \\
  \hline
\end{tabular}
\end{center}


\small{

\end{document}